\documentclass[%
reprint,
aps,
]{revtex4-1}

\usepackage{graphicx}

\usepackage{dcolumn}
\usepackage{bm}
\usepackage{siunitx}

\usepackage{amsmath}
\usepackage{amssymb}
\usepackage[makeroom]{cancel}

\usepackage{float}

\usepackage{tabularx}

\usepackage{hyperref}
\usepackage{xcolor}
\hypersetup{frenchlinks = true,
	colorlinks,
	linkcolor={red!50!black},
	citecolor={blue!50!black},
}

\usepackage{IEEEtrantools}

\usepackage{braket}
\usepackage{dsfont}

\def\st#1{_{\text{#1}}}

\def\op#1{\hat{#1}}

\newcommand{\au}{\uparrow}
\newcommand{\ad}{\downarrow}
\newcommand{\sqswap}{$\sqrt{\text{SWAP}}$ }
\renewcommand{\d}[0]{\mathrm{d}}

\usepackage{todonotes}

\usepackage{tikz}
\usetikzlibrary{arrows, decorations.pathmorphing}

\renewcommand{\vec}[1]{\boldsymbol{#1}}

\begin{document}

	\title{Time-Optimal Control of Collisional \sqswap Gates in Ultracold Atomic Systems}
	
	\author{Jesper Hasseriis Mohr Jensen}
	\author{Jens Jakob S\o rensen}
	\author{Klaus M\o lmer}	
	\author{Jacob Friis Sherson}	
	
	\email{sherson@phys.au.dk}
	\affiliation{%
		Department of Physics and Astronomy, Aarhus University, Ny Munkegade 120, 8000 Aarhus C, Denmark
	}%

	\date{\today}

	\begin{abstract}
		We use quantum optimal control to identify fast collision-based two-qubit \sqswap gates in ultracold atoms. We show that a significant speed up can be achieved by optimizing the full gate instead of separately optimizing the merge-wait-separate sequence of the trapping potentials. Our optimal strategy does not rely on the atoms populating the lowest eigenstates of the merged potential, and it crucially includes accumulation of quantum phases before the potentials are fully merged. Our analyses transcend the particular trapping geometry, but	to compare with previous works, we present systematic results for an optical lattice and find greatly improved gate durations and fidelities.

	\end{abstract}
	
	\maketitle
	
\section{\label{sec:sqswap} Introduction}

Optically trapped ultracold atomic systems have enjoyed impressive recent progress with regards to their preparation and control of both internal and external degrees of freedom \cite{chu2002cold,bloch2008many,sherson2010single,weitenberg2011single}.
Especially, recent
advances  \cite{kaufman2014two,wang2015coherent,kim2016situ,endres2016atom,barredo2016atom,lee2016three,kumar2018sorting,barredo2018_3dassembly,saskin2019narrow,norcia2018microscopic,cooper2018alkaline} have augmented the viability of using the long coherence times of their spin internal degrees of freedoms for quantum computing \cite{brennen1999quantum,divincenzo2000physical,jaksch2000fast,lukin2001dipole,daley2008quantum,negretti2011quantum,weitenberg2011quantum,schneider2012quantum,byg2014Superlattice,pagano2019fast}.
While single-qubit operations with fidelities above $0.99$ have been demonstrated in multiple experiments \cite{xia2015randomized,wang2016single},
corresponding fidelities of two-qubit entangling operations are still subject of research  \cite{mandel2003controlled,anderlini2007controlled,wilk2010entanglement,zhang2010deterministic,isenhower2010demonstration,maller2015rydberg,kaufman2015entangling,jau2016entangling,levine2018high}.
Entangling two-qubit gates can be mediated by long-range interactions 
such as  dipole-dipole interactions between Rydberg atoms \cite{jaksch2000fast,browaeys2016experimental,saffman2010quantum,saffman2016quantum}.
Although the long-range nature of these interactions allows potentially fast operations, their use of highly excited atomic states make them vulnerable to enhanced coupling to the environment.

\begin{figure}[t]
	\includegraphics[width=\linewidth]{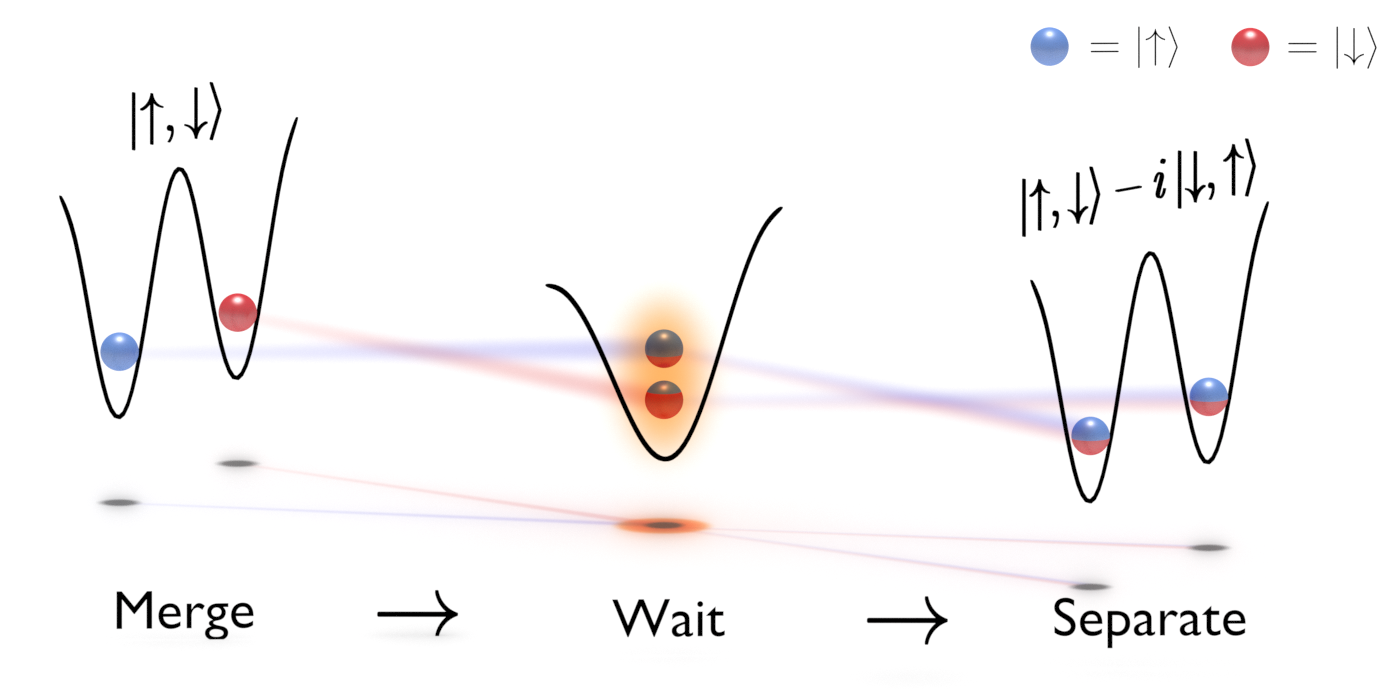}
	\caption
	{
		(Color online) Schematic illustration of the merge-wait-separate sequence implementing the \sqswap gate. Completing the sequence transfers a state of initially opposite spins into a spin-entangled state.
	}
	\label{fig:spinexchange}
\end{figure}

Short-range collisional (contact) interactions provide an alternative for neutral atom quantum gates \cite{jaksch1999entanglement,calarco2000quantum,anderlini2007controlled,hayes2007quantumlogic,kaufman2015entangling}.
Merging two initially separated atoms into a common trap initiates a collisional interaction depending on the exchange symmetry of the atomic wave function and hence of the spin state of the atoms.
After a duration determined by the interaction strength in the merged state, the atoms are spatially separated, and
under appropriate conditions, the simple three-stage merge-wait-separate sequence illustrated in Fig.~\ref{fig:spinexchange} realizes the entangling \sqswap gate. 

The short-range character of the collisions ensures that only the desired qubits participate in the operation, decreasing the detrimental coupling to other qubits and to the environment. Relying on collisional interaction imposes strong requirements for the precision with which the spatial degree of freedom of the atoms must be controlled. Current experimental control protocols to perform this entangling gate are thus adiabatic in nature, but this severely limits the total number of gate operations before decoherence effects become significant \cite{anderlini2007controlled,kaufman2015entangling}.

\begin{center}
	\begin{figure*}[t]
		\includegraphics[]{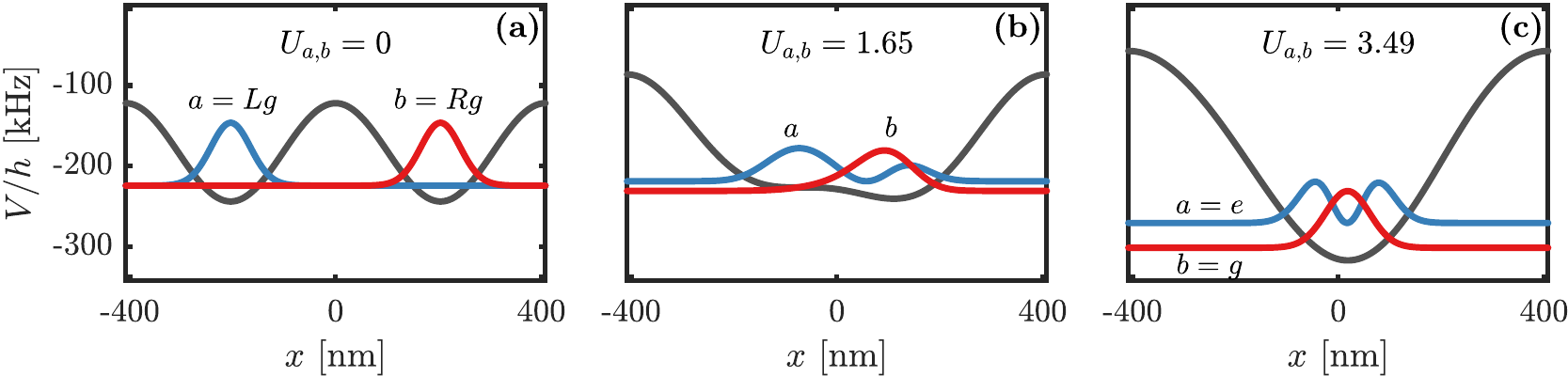}
		\caption{
			(Color online)
			Merging stage ($\beta=0.52\pi \times  t/T$, $\theta=-0.474\pi$, $V_0/h=122\,\si{kHz}$) of lattice unit cell in the independent-particle picture.
			\textbf{(a)} $t=0:$ Atoms are initially prepared in the separated double-well configuration.
			\textbf{(b)} $t=0.7T:$ Intermediate snapshot of the merging process.
			\textbf{(c)} $t=T:$ Atoms occupy orthogonal states in the merged single-well configuration.
			In each snapshot the energy difference $U_{a,b}$ between corresponding $\ket{\Psi^\pm_{a,b}}$ is shown in units of $\si{kHz}\cdot h$  (see Eqs.~\eqref{eq:spstates}-\eqref{eq:Uab}).
		}
		\label{fig:VLattice}
		\vspace{-0.19cm}
	\end{figure*}
\end{center}

\vspace{-0.35cm}
Finding fast, complex control protocols is a task well-suited for quantum optimal control.
Previous works \cite{de2008optimal,mundt2009optimalcontrol} have thus 
reduced the duration of the merging stage in optical geometries by orders of magnitude compared to adiabatic solutions. However, the current best results for the merging stage have thus far not crossed the 0.99 threshold. This is especially detrimental since even slight merging  errors also reduce the quality of subsequent waiting and separation stages. To our knowledge, no optimization of the full gate has been carried out.

In this paper, we discuss how the \sqswap gate relies on the evolution of a relative phase between singlet and triplet spin state components, and why the partial accumulation of this phase already during the merging stage is a challenge for the optimization of stagewise protocols. We develop means to solve this challenge, and we proceed to show that a protocol without dedicated merge-wait-separate stages yields faster performance and 0.99 fidelity for the full \sqswap operation with ultracold $^{87}$Rb atoms.
We stress that our considerations of the accumulated relative phase and the full gate optimization are independent of the specific physical problem geometry, atomic species as well as model dimensionality.

The paper is organized as follows.
In Sec.~\ref{sec:geometries} we present the trapping geometry under consideration for the \sqswap operation.
In Sec.~\ref{sec:physical_model} we present the theory for implementing the \sqswap gate in ultracold atoms. The two-particle Hamiltonian is introduced and general properties of the allowed states are discussed.
Initially, we use symmetrized product states to construct the computational basis states. Using these states, we describe the important accumulated relative phase during the merging sequence. In our numerical optimization, we do not rely on the independent particle approximation but propagate genuine two-particle wave functions for the interacting atoms.
In Sec.~\ref{sec:qoc} we discuss the difference between the staged merge-wait-separate approach and the full gate approach in terms of optimal trajectories in Hilbert space.
In Sec.~\ref{sec:results} we present and discuss the results.
In Sec.~\ref{sec:conclusions} we summarize the main conclusions of the paper.

\section{\label{sec:geometries}  Trapping geometry}
A necessary feature of any candidate geometry for the collisional \sqswap operation is the possibility to bring two atoms from a separated configuration into contact by, for example, merging them in a common trap as illustrated in Fig.~\ref{fig:spinexchange}.
We consider the implementation with an optical lattice \cite{anderlini2006controlled,anderlini2007controlled} that has been loaded with a Mott state of unit filling and where the atoms at every other site have been prepared with opposite spin states.
An analysis presented in Appendix~\ref{app:1ddesc} justifies a 1D description with the potential

	\begin{align}
	V(x) = -V_0\bigg[&\cos^2\left(\frac{\beta}{2}\right)\left\{1+\cos^2(k x-\frac{\pi}{2})\right\} \nonumber \\
	 + &\sin^2\left(\frac{\beta}{2}\right)\left\{1+\cos(k x -\theta-\frac{\pi}{2})\right\}^2\bigg]. \label{eq:VLattice}
	\end{align}
Here $V_0$ provides  the overall lattice depth, while $\beta$ and $\theta$ adjust the height and tilt of adjacent wells.
By controlling  $\{\beta(t),\theta(t),V_0(t)\}$, pairs of adjacent wells are transformed from a double-well into a single-well configuration as illustrated in Fig.~\ref{fig:VLattice}.
In an independent-particle picture, the atom in the \textbf{g}round state of the \textbf{L}eft (\textbf{R}ight) well is transferred to the first \textbf{e}xcited (\textbf{g}round) state of the merged well, $\ket{Lg}\ket{Rg} \rightarrow \ket{e}\ket{g}$.

\section{\label{sec:physical_model}  \sqswap gate with cold Atoms}
The \sqswap gate is concerned with the qubit, i.e., spin degrees of freedom of the atoms, and its simplest implementation is through the three-stage merge-wait-separate sequence in Fig.~\ref{fig:spinexchange}.
In this section, we recall the theory behind this procedure.

The system is described by the effective Hamiltonian
\begin{align}
\op H
&= \sum_{i=1}^2 \op  h(x_i) + g_\text{1D}\delta(x_1-x_2).
\label{eq:H}
\end{align}
Here, $x_i$ are the coordinates of the two atoms, $\op  h(x) = -\frac{\hbar^2}{2m}\frac{\partial^2}{\partial x_i^2} + V(x)$ is the single-particle Hamiltonian with trapping potential $V$. The form of the interaction term and the value of the coupling strength $g_\text{1D}$ are discussed in Appendix~\ref{app:1ddesc}.

We use identical bosonic atoms and the full two-particle state can thus be expanded on symmetric states of the form
\begin{align}
\ket{\Phi^{\au\au}} &= \ket{\Psi^+} \ket {\chi^{\au\au}},
&&\ket{\Phi^{\ad\ad}} = \ket{\Psi^+} \ket {\chi^{\ad\ad}}, \label{eq:fullstateothers}\\ 
\ket{\Phi^\pm} & = \ket{\Psi^\pm} \ket {\chi^\pm}, \label{eq:fullstate}
\end{align}
with the symmetric ($+$) spin triplet and anti-symmetric $(-)$ spin singlet states
\begin{align}
\ket {\chi^\pm}  = \ket{\au}_1\ket{\ad}_2 \pm \ket{\ad}_1\ket{\au}_2,
\end{align}
and symmetric ($+$) and anti-symmetric $(-)$ spatial states $\ket{\Psi^\pm}$,
where proper normalization is implied in the remainder of the paper.

The Hamiltonian \eqref{eq:H} is independent of the spin degrees of freedom and cannot induce transitions between states of different total spin. 
The interaction term $g_\text{1D}\delta(x_1-x_2)$ only acts on the $\Psi^+$ component since  $\Psi^-(x_1,x_2=x_1) = 0$ by construction.
This introduces an energy difference between the otherwise degenerate singlet and the triplet state vector components. This energy difference is the main mechanism behind the \sqswap gate.

\subsection{Symmetrized Product States}

To illustrate the dynamics leading to the \sqswap operation we consider an approximate analysis with symmetrized product states, but we emphasize that our numerical optimization is carried out with the full two-particle interaction dynamics. 

We associate the qubits with the spins of atoms occupying definite spatial states. If $\ket a$ and $\ket b$ denote such single-particle eigenstates of $\op h(x)$,
spatially symmetrized product states 

\begin{align}
\ket{\Psi_{a,b}^\pm} &=\ket{a}_1\ket b_2 \pm  \ket b_1 \ket a_2.
\label{eq:spstates}
\end{align}
are approximate eigenstates of $\op H$ with energies 
\begin{align}
\braket{\Psi_{a,b}^- | \op H |\Psi_{a,b}^-} &= E_a + E_b, \\
\braket{\Psi_{a,b}^+ | \op H |\Psi_{a,b}^+} &= E_a + E_b + U_{a,b},
\end{align}
where $E_a$ and $E_b$ are single-particle energies and
\begin{align}
U_{a,b} &\equiv 2 \int_{-\infty}^{\infty} |a(x)|^2|b(x)|^2 g_{\text{1D}}(x) \d x.
\label{eq:Uab}
\end{align}
The energy difference between $\ket{\Psi_{a,b}^\pm}$ clearly depends on the spatial overlap of the two atoms and the collisional coupling strength.

Using \eqref{eq:spstates} in \eqref{eq:fullstateothers}-\eqref{eq:fullstate}, we define symmetrized computational basis states for the \sqswap operation: 
\begin{align}
\hspace{-0.3cm}\ket{\au_a,\ad_b} &\equiv \ket{\au_a}_1\ket{\ad_b}_2 + \ket{\ad_b}_1\ket{\au_a}_2 \hspace{-0.3cm}&&=\ket{\Phi^+_{a,b}} + \ket{\Phi^-_{a,b}},\label{eq:updown} \\
\hspace{-0.3cm}\ket{\ad_a,\au_b} &\equiv \ket{\ad_a}_1\ket{\au_b}_2 + \ket{\au_b}_1\ket{\ad_a}_2 \hspace{-0.3cm}&&=\ket{\Phi^+_{a,b}} - \ket{\Phi^-_{a,b}} , \label{eq:downup}  \\
\hspace{-0.3cm}\ket{\au_a,\au_b} &\equiv \ket{\au_a}_1\ket{\au_b}_2 + \ket{\au_b}_1\ket{\au_a}_2 \hspace{-0.3cm}&&=\ket{\Phi^{\au\au}_{a,b}}, \label{eq:upup}  \\
\hspace{-0.3cm}\ket{\ad_a,\ad_b} &\equiv \ket{\ad_a}_1\ket{\ad_b}_2 + \ket{\ad_b}_1\ket{\ad_a}_2 \hspace{-0.3cm}&&=\ket{\Phi^{\ad\ad}_{a,b}}. \label{eq:downdown}
\end{align}
We see in the two first lines that the relative phase between the triplet and singlet components is essential to determine how the spins are correlated with the spatial states of the atoms.
The two last lines represent states that are unaffected by the \sqswap operation and acquire only the same phase factor as the $(+)$ components in the two first lines.

We define the \sqswap gate in the states \eqref{eq:updown}-\eqref{eq:downdown} as
\begin{align}
\hspace{-0.2cm}\ket{\au_a,\ad_b} \rightarrow \sqrt{\text{SWAP}}\ket{\au_a,\ad_b}  &=  \hspace{0.33cm}\ket{\au_{a},\ad_{b}} -i\ket{\ad_{a},\au_{b}}, \label{eq:sqswapbasis} \\
\hspace{-0.2cm}\ket{\ad_a,\au_b} \rightarrow \sqrt{\text{SWAP}}\ket{\ad_a,\au_b}  &= -\ket{\au_{a},\ad_{b}} - i\ket{\ad_{a},\au_{b}}, \label{eq:sqswapbasis2}\\
\hspace{-0.2cm}\ket{\au_a,\au_b} \rightarrow \sqrt{\text{SWAP}}\ket{\au_a,\au_b} &= e^{-i\pi/4}\ket{\au_a,\au_b}, \label{eq:sqswapbasis3}\\
\hspace{-0.2cm}\ket{\ad_a,\ad_b} \rightarrow \sqrt{\text{SWAP}}\ket{\ad_a,\ad_b} &= e^{-i\pi/4}\ket{\ad_a,\ad_b}. \label{eq:sqswapbasis4}
\end{align}
If the spins are initially in opposite states, the \sqswap operation yields an entangled state. If the spins are initially equal a phase $e^{-i\pi/4}$ is applied. In the following, we focus on the mapping \eqref{eq:sqswapbasis} and study the approximate system dynamics to find the necessary conditions for implementing the \sqswap gate.

\subsubsection*{Time Evolution in Static Trap}

\begin{figure}[t]
	\begin{center}
		\centering
		\includegraphics[]{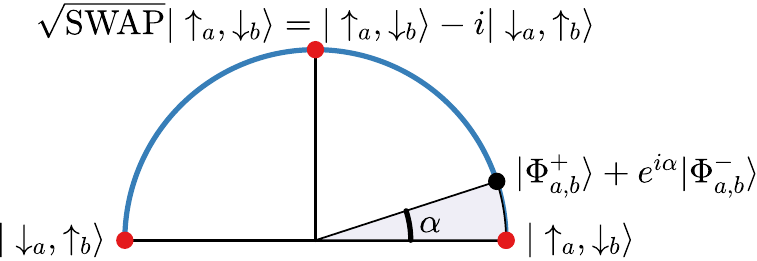} 	
		\caption{(Color online) Illustration of how the spins are distributed due to the relative phase in the first two quadrants. Here $\alpha = \pi/10$.
		}
		\label{fig:phaseplane}	
		\includegraphics[]{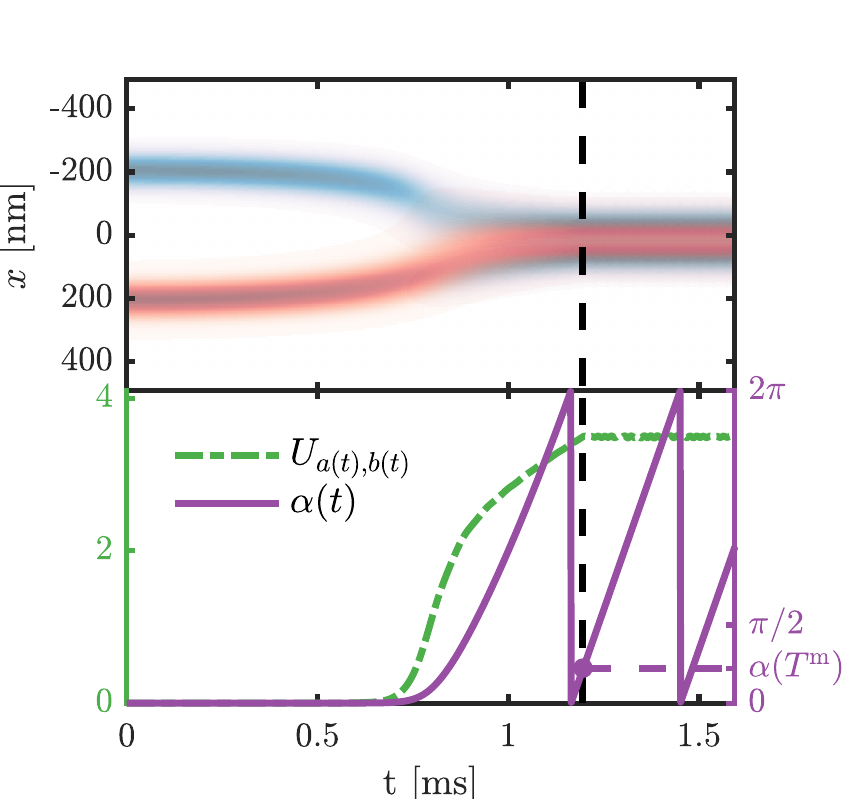}
	\end{center}
	\caption{(Color online) Adiabatic lattice merging of the atoms into single-well configuration as in Fig.~\ref{fig:VLattice}.
		Top: Density plots $|a(x,t)|^2$ (blue) and $|b(x,t)|^2$ (red) of single-particle states $\ket{a(0)} \ket{b(0)}=\ket{Lg}\ket{Rg}\rightarrow \ket{e}\ket{g}$. The merge duration $T^{\mathrm{m}}$ 
		is marked by the dashed line and at $t>T^{\mathrm{m}}$ the potential is static.
		Bottom: Interaction energy $U_{a(t),b(t)}$ Eq.~\eqref{eq:Uab} (green dash-dotted) and total accumulated relative phase $\alpha(t) =\hbar^{-1} \int_0^t U_{a(t'),b(t')} \d t'$ (purple solid).
		The phase acquired during merging $\alpha(T^{\mathrm{m}})$ is non-zero.
	}
	\label{fig:adiabaticMerge2p}
	
\end{figure}

If $\ket a$ and $\ket b$ are single-particle eigenstates of $\op h(x)$,
the time evolution of $\nobreak\ket{\au_a,\ad_b}  \nobreak= \nobreak\ket{\Phi^+_{a,b}} + \ket{\Phi^-_{a,b}}$ is approximately given by
\begin{align}
\ket{\Phi_{a,b}(t)} & \approx  \ket{\Phi^+_{a,b}} +  e^{i\alpha}\ket{\Phi^-_{a,b}} \label{eq:swapphi}\\
&\rightarrow \cos\left(\frac{\alpha}{2}\right) \ket{\au_a,\ad_b} - i \sin\left(\frac{\alpha}{2} \right)\ket{\ad_a,\au_b} \label{eq:swap}
\end{align}
where we disregard global phases as the dynamics of the system and the spin distribution on the atoms is fully described by the relative phase $\alpha(t) = U_{a,b} t/\hbar$ between the state components.
After the duration $T_{\text{SWAP}} = \pi\hbar/U_{a,b}$  ($\alpha = \pi$) the spins are fully swapped $\ket{\ad_a,\au_b}$ while the interaction for half of this duration $T_{\sqrt{\text{SWAP}}} = T_{\text{SWAP}}/2$ ($\alpha = \pi/2$) implements the desired \sqswap gate as illustrated in Fig.~\ref{fig:phaseplane}.

\medskip
The spin swapping rate $\dot \alpha \propto U_{a,b}$ depends only on the interaction energy Eq.~\eqref{eq:Uab}, which remains constant throughout the evolution in a static trap.
To obtain a finite $T_{\sqrt{\text{SWAP}}}$ the atoms must therefore be sufficiently overlapping. This condition is satisfied in the merged configuration,  but clearly not in the initial configuration
-- see Fig.~\ref{fig:VLattice}\textbf{(a)} and \textbf{(c)}.

\subsubsection*{Time Evolution in Adiabatically Transformed Trap}
Suppose the trapping potential is transformed adiabatically with respect to the single-particle states such that $a=a(t)$ and $b=b(t)$ follow instantaneous eigenstates of $\op h(x)$. Then the time evolution given by Eq.~\eqref{eq:swap} remains valid with $\alpha(t) =\hbar^{-1} \int_0^t U_{a(t'),b(t')} \d t'$. The spin swapping rate $\dot \alpha \propto U_{a(t),b(t)}$ is now time-dependent and becomes non-zero as the atoms begin to overlap.

\medskip

As a consequence, the phase accumulated during the merging is in general $\alpha(T) \neq 0$, which reduces the waiting stage duration needed to obtain the entangled state. 
This is clearly seen in Fig.~\ref{fig:phaseplane} if we let $\alpha=\alpha(T)$; the phase acquired during merging already brought the state closer to $\sqrt{\text{SWAP}}\ket{\au_a,\ad_b}$. An additional $\alpha(T)$ will be acquired during the separation stage.
Assuming for simplicity $\alpha(T) \leq \pi/4$, the \sqswap duration is reduced to $T_{\sqrt{\text{SWAP}}} = (\pi/2- 2\alpha(T))\hbar/U_{a(T),b(T)}$ where the factor 2 accounts for both the merging and separation stage.
Labeling explicitly the merging duration by $T\rightarrow T^{\mathrm{m}}$, the total duration of the full gate operation using the merge-wait-separate sequence is then $T^{\mathrm{f}}=2 T^{\mathrm{m}} + T_{\sqrt{\text{SWAP}}}$.

This sequence explicitly implements the mapping of a single basis state \eqref{eq:sqswapbasis}. Fortunately, such sequence also simultaneously realizes the remaining mappings.
The mapping \eqref{eq:sqswapbasis2} simply corresponds to starting at $\alpha(0)=\pi$ from where the phase accumulation proceeds identically to above.
For the mappings \eqref{eq:sqswapbasis3}-\eqref{eq:sqswapbasis4}, note that \eqref{eq:sqswapbasis} implies correct, simultaneous preparation of the individual triplet and singlet components since there is no coupling between states of different symmetry. Thus, these mappings are also guaranteed to be realized.
We note at this point that the converse is not necessarily true: a sequence implementing mappings \eqref{eq:sqswapbasis3}-\eqref{eq:sqswapbasis4} does not guarantee the mappings \eqref{eq:sqswapbasis}-\eqref{eq:sqswapbasis2} since the singlet component is absent.

To summarize the ideas developed in this section, we show a numerical example of the accumulated phase to illustrate qualitative features in Fig.~\ref{fig:adiabaticMerge2p}.
Here, the system from Fig.~\ref{fig:VLattice} is adiabatically merged and followed by a holding time in the static final potential \footnote{Visit \url{https://www.quatomic.com/quatomic_publications/} for animations of the single-particle densities.}.
The single-particle states are propagated independently with the interaction only affecting the relative phase and not the spatial distribution, which is the approximation made in Eq.~\eqref{eq:swapphi}.
At each point in time we construct $\ket{\Psi_{a(t),b(t)}^\pm}$ and calculate the corresponding $U_{a(t),b(t)}$.
As the atoms begin to overlap the relative phase accumulates resulting in $\alpha(T^{\mathrm{m}}) \approx \pi/4$.
Following the merge, the $
\sqrt{\text{SWAP}}\ket{\au_e,\ad_g}$ state is obtained after a short, static holding time of about $0.04\,\si{ms}$.
If instead the atoms were immediately separated, the $\sqrt{\text{SWAP}}\ket{\au_{Lg},\ad_{Rg}}$ state would be obtained since  $2 \alpha(T^{\mathrm{m}}) \approx \pi/2$.

The adiabatic transfer thus ensures a high-fidelity implementation of the \sqswap gate since we are guaranteed to stay within the superposition of just one singlet and triplet state Eq.~\eqref{eq:swapphi}.
However, for the purposes of quantum computation we also need the implementation to be fast.
The speed up is achieved by exploiting the interference effects of many intermediately populated excited states. To enable the engineering of these very complicated diabatic transfers we turn to quantum optimal control. Before doing so, we close this section by replacing the symmetrized product states with the true two-particle eigenstates.

\subsection{Two-Particle Eigenstates}
The analysis of the dynamics in the previous section approximated the symmetrized product states $\ket{\Psi_{a,b}^\pm}$ Eq.~\eqref{eq:spstates} to be eigenstates of $\op H(x_1,x_2)$ where $\ket a$ and $\ket b$ were eigenstates of $\op h(x)$.
In the limit of vanishing interactions (no spatial overlap or zero coupling) this approximation is exact. This allows us to relate to the true spatial two-particle eigenstates (annotated with $\sim$) in the following way:
\begin{align}
\ket{\tilde \Psi_{a,b}^-} &=\, \ket{\Psi_{a,b}^-}, \\
\ket{\tilde \Psi_{a,b}^+} &\rightarrow \ket{\Psi_{a,b}^+},
\end{align}
where $\rightarrow$ in this context implies vanishing interactions.
Only the triplet state is affected by the interaction as the system can lower its energy by depleting the diagonal $x_1=x_2$.
This notation is very convenient since we retain reference to the intuitive independent-particle picture. In particular, the analysis following Eq.~\eqref{eq:spstates} of the previous section is still valid upon annotating all states and energies with '$\sim$'.
The approximation made in Eq.~\eqref{eq:swapphi} consisted in ignoring the small interaction matrix elements between different triplet states (other off-diagonal elements vanish identically) such that the symmetrized product states $\ket{\Psi^\pm_{a,b}}$ were approximate eigenstates of the full Hamiltonian.
In using the true eigenstates $\ket{\tilde \Psi^\pm_{a,b}}$, the results are exact.

It is also possible to describe the dynamics in a static potential with an effective spin Hamiltonian $\op H_\text{spin}= J_{\text{ex}}\cdot\op{\vec S}_1 \otimes \op{\vec S}_2$ even though the spin swapping is purely due to spatial effects (extension to the adiabatic case is straightforward). This is also known as the exchange interaction. Here $\op{\vec S}_i$ are spin operators and $J_{\text{ex}} = U_{a,b}$ is the exchange energy. See Appendix \ref{appendix:spinham} for a brief derivation of this result.

\bigskip
\section{Quantum Optimal Control of \sqswap in Ultracold Atoms } \label{sec:qoc}
In Sec.~\ref{sec:physical_model} we showed that the desired \sqswap operation can be implemented based on a single basis state mapping 
$\ket{\au_{Lg},\ad_{Rg}} \rightarrow \sqrt{\text{SWAP}}\ket{\au_{Lg},\ad_{Rg}}$. Formulating this as a state transfer control problem, the initial- and target states for the full gate are
\begin{align}
\ket{\Phi_0} &= \ket{\tilde \Phi^+_{Lg,Rg}} +  \ket{\tilde \Phi^-_{Lg,Rg}}, \label{eq:initialstate}\\
\ket{\Phi^{\text{f}}_\text{t}} &=   \ket{\tilde \Phi^+_{Lg,Rg}} + e^{i\pi/2}\ket{\tilde \Phi^-_{Lg,Rg}}. \label{eq:ftarget}
\end{align}
By writing the states in terms of singlet and triplet components we emphasize the goal of ultimately establishing the correct relative phase in the separated configuration by letting the atoms collide. See Appendix~\ref{app:methods} for a description of methods and problem parameters.

A simplified approach to solving the control problem $\ket{\Phi_0} \rightarrow \ket{\Phi^{\text{f}}_\text{t}}$ is to use the merge-wait-separate sequence.
The problem can then be reduced to optimizing just the merging stage as the associated optimal controls can be extended to implement the whole sequence:
the waiting stage duration is determined by the relative phase $\alpha(T^\mathrm{m})$ acquired during the optimized merging stage, whereas the separation stage is carried out by propagating along the time-inverted optimized merging control.
A suitable target state for the merging sub-problem is
\begin{align}
\ket{\Phi^{\text{m}}_\text{t}} &=   \ket{\tilde \Phi^+_{e,g}} +  e^{i\alpha_\text{t}}\ket{\tilde \Phi^-_{e,g}}. \label{eq:mtarget}
\end{align}
The merging sub-problem thus consists in realizing $\ket{\Phi_0} \rightarrow \ket{\Phi^{\text{m}}_\text{t}}$.
 This target state is not stationary and will exhibit two-level beating dynamics if the transfer is successful.
Note the inclusion of a target relative phase $\alpha_\text{t} \leq \pi/4$.
This is because $\alpha(t)$ is monotonically increasing due to $\tilde U\geq 0$. If the target phase is excluded ($\alpha_\text{t}=0$), then, in an independent-particle picture, the optimizer will
try to minimize the time-integrated overlap between the atomic states during the transfer. This is contradictory to the overall goal of the merging, which is exactly to overlap the atoms to enable the spin swapping.
The total accumulated phase is not crucial to the overall \sqswap operation since one may simply adjust the duration of the waiting stage.
It follows that any final superposition of $\ket{\tilde \Phi^\pm_{e,g}}$ is acceptable as long as $\alpha(T) \leq \pi/4$.   In fact, a non-zero phase within this range is beneficial as it speeds up the overall gate operation, see Fig.~\ref{fig:phaseplane}. However, the standard figure of merit for the state transfer quality is the fidelity
\begin{align}
\mathcal F =|\braket{\Phi_\text{t} | \Phi(T)}|^2,
\label{eq:f}
\end{align}
which for the merging sub-problem, $\ket{\Phi_\text{t}} = \ket{\Phi^{\text{m}}_\text{t}}$, depends on the chosen target phase.
In this case a more appropriate measure of the transfer quality insensitive to the relative phase
is the total population in $\ket{\tilde \Phi^\pm_{e,g}}$:
\begin{align}
\mathcal{F}'&=
|\braket{\tilde \Phi^+_{e,g}| \Phi(T)}|^2 + |\braket{\tilde \Phi^-_{e,g}| \Phi(T)}|^2 \geq \mathcal{F}, \label{eq:f'}
\end{align}
where  $\mathcal F=\mathcal F'$ only when $\alpha(T)=\alpha_\text{t}$. Thus, optimizing $\mathcal F'$ would alleviate the constraint on the relative phase. Instead of doing this, we simply use $\mathcal{F}'$ as a stopping condition and optimize $\mathcal F$ with an appropriate target phase such that $\mathcal F\approx \mathcal F'$.
For the durations $T=T^\mathrm{m}$ under consideration, numerical investigations suggest typical values of $\alpha(T^\mathrm{m}) \in [0.31,0.44] \approx [\pi/10, \pi/7]$. This is well below $\pi/4$.
We find a suitable target phase to be $\alpha_\text{t} = 0.33$.
The optimization could be improved by making an appropriate cost functional replacement $J_{\mathcal{F}} \rightarrow J_{\mathcal{F'}}$ and deriving the resulting optimality system.
\begin{figure}[t]
	\includegraphics[]{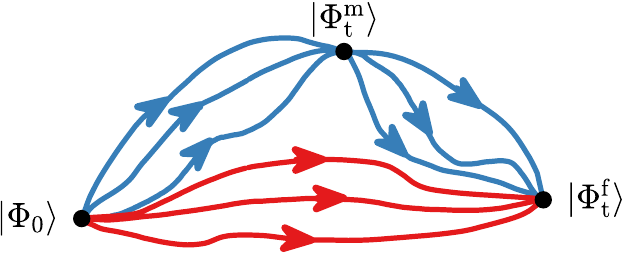}
	\caption{(Color online) Schematic illustration of different idealized optimal trajectories in Hilbert space, both leading to   $\sqrt{\text{SWAP}}\ket{\au_{Lg},\ad_{Rg}}$. The upper trajectories (blue) corresponds to implementing the simple merge-wait-separate sequence. In this case, each trajectory must pass through $\ket{\Phi^{\text{m}}_\text{t}}$ from where it can be connected to $\ket{\Phi^{\text{f}}_\text{t}}$ as explained in the main text. The lower trajectories correspond to implementing the full gate without distinct stages. Here there is no requirement to pass through a particular intermediate state.
	}
	\label{fig:trajectory}
\end{figure}

Previous works \cite{de2008optimal,mundt2009optimalcontrol} considered only the merging sub-problem and did not include the singlet component in initial nor target state, using only the triplet component.
This corresponds to realizing the mappings \eqref{eq:sqswapbasis3}-\eqref{eq:sqswapbasis4}, which as previously mentioned does not guarantee simultaneous implementation of the remaining mappings \eqref{eq:sqswapbasis}-\eqref{eq:sqswapbasis2} but may still be considered an approximation.
Additionally, ignoring the singlet component reduces the problem difficulty since the optimization no longer has to achieve a relative phase.

The merit of solving the full control problem by reducing it to the merging sub-problem is its conceptual simplicity.
From a numerical point of view, it also typically involves a reduced interval of time integration.
This is substantial due to the low time resolution required to faithfully simulate the interaction $\delta$-function.
Nevertheless, there are several drawbacks to this approach.
Firstly, optimizing towards $\ket{\Phi^{\text{m}}_\text{t}}$ is an artificial and unnecessarily strict condition.
It can be understood as forcing the optimal state trajectories to pass through a particular intermediate point (or small volume in case of $\mathcal F'$) in Hilbert space as illustrated by the upper trajectories in Fig.~\ref{fig:trajectory}.
Secondly, extending optimal controls to implement the complete merge-wait-separate sequence is predicated on idealized unit fidelity transfers wrt. $\ket{\Phi^{\text{m}}_\text{t}}$. Even 0.99 fidelity solutions will have their errors exacerbated throughout the waiting- and separation stages, causing alterations to the state trajectory away from $\ket{\Phi^{\text{f}}_\text{t}}$, which is the state we are ultimately interested in obtaining.
The lower trajectories in Fig.~\ref{fig:trajectory} corresponds to optimizing the full problem $\ket{\Phi_0} \rightarrow \ket{\Phi^{\text{f}}_\text{t}}$ directly and is not required to pass through any particular intermediate state. The control problem is not broken up into distinct stages and becomes much less restrictive.

In this work we combine the two approaches. We first optimize towards $\ket{\Phi^{\text{m}}_\text{t}}$ in the merging sub-problem. We subsequently extend the corresponding optimized controls to implement the full gate. These extended controls are then used as seeds for the optimization towards $\ket{\Phi^{\text{f}}_\text{t}}$ in the full gate problem. This methodology allows a fair comparison of the two approaches.

\section{Results}
\label{sec:results}

\begin{figure*}[t]
	\begin{center}
	\begin{minipage}{.483\textwidth}
	\includegraphics[]{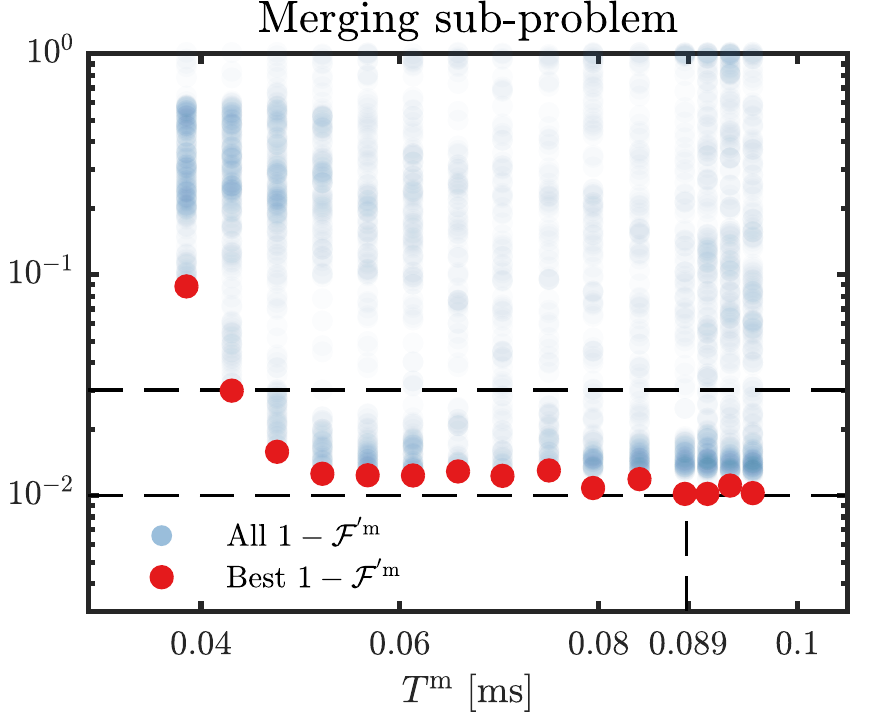}
	\caption{(Color online) Optimization results (lower is better) for the merging sub-problem.
		The lower (upper) horizontal dashed line marks the 0.99 (0.97) fidelity threshold.
			$1-\mathcal F^{'\mathrm{m}}$ is shown for each solution in a batch optimized for $\alpha_\text{t}=0.33$. The distribution density is indicated by the translucency. The best solutions for each $T$ are marked in solid red. Out of 2544 seeds, only 3 optimized to $\mathcal F^{'\mathrm{m}}=0.99$.
		The quantum speed limit bound in this optimization batch is $T^{\mathrm{m}}_\text{QSL} \leq 0.0888 \,\si{ms}$.
	}
	\label{fig:FT_dechiara}
	\end{minipage}
	\hspace{0.4cm}
	\begin{minipage}{.483\textwidth}
	\includegraphics[]{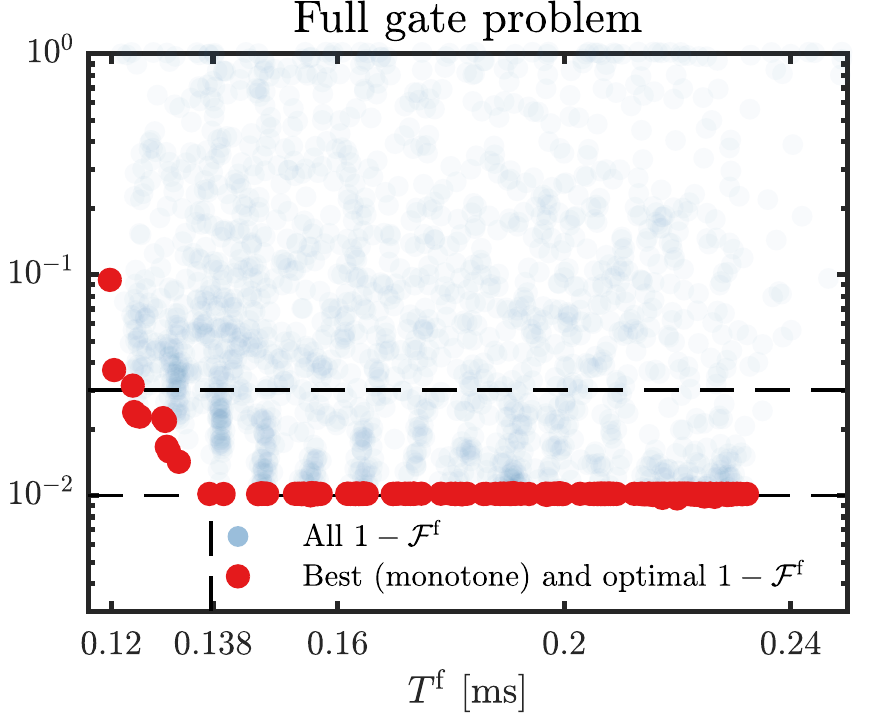}
	\caption{(Color online) Optimization results (lower is better) for the full gate problem when using the optimized extended solutions from Fig.~\ref{fig:FT_dechiara} as seeds.
	The lower (upper) horizontal dashed line marks the 0.99 (0.97) fidelity threshold.
		$1-\mathcal F^{\mathrm{f}}$ is shown for each solution.
		The monotonically best and optimal solutions are marked in solid red.
		Out of 2323 seeds, a total of 277 optimized to $\mathcal F^{\mathrm{f}}=0.99$.
		The quantum speed limit bound in this optimization batch is $T^{\mathrm{f}}_\text{QSL} \leq 0.1377 \,\si{ms}$.
	}
	\label{fig:FT_dechiara_full}
\end{minipage}
	\end{center}

\end{figure*}

In this section we present optimization results.
A solution is considered optimal if it exceeds $ \mathcal{F}^{'\text{m}} = 0.99$ in the sub-merging problem or $ \mathcal{F}^{\text{f}} = 0.99$ in the full problem.

\begin{figure*}[h]
	\includegraphics[scale=0.98]{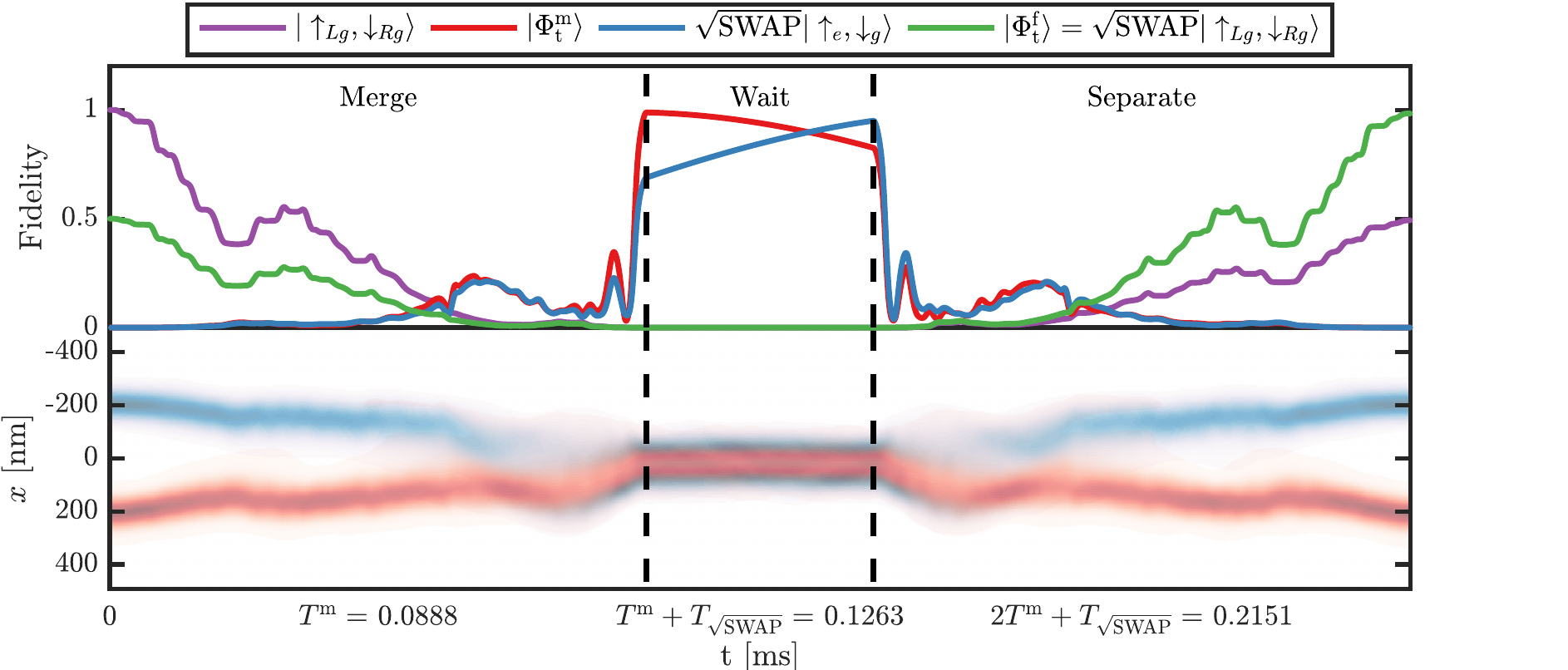}
	\caption{Full \sqswap gate operation based on the time-optimal merging control. The merge-wait-separate stages are indicated with the vertical dashed lines.
		 Top: Instantaneous fidelities with various states. Bottom: Corresponding independent single-particle densities. The target state $\ket{\Phi^{\mathrm{m}}_\mathrm{t}}$ is obtained at $T^{\mathrm{m}}$ with $\mathcal F^{'\text{m}}=0.99$.  The system then exhibits two-level dynamics before it is separated into the initial configuration achieving $\ket{\Phi^{\mathrm{f}}_\mathrm{t}}$ with sub-optimal final fidelity $\mathcal F^{\text{f}}=0.983$.}
	\label{fig:fullcontrolpropagation}
\end{figure*}

Fig.~\ref{fig:FT_dechiara} shows optimization results for the merging sub-problem. A total of 2544 seeds were optimized with between 100 and 250 seeds per duration.
The red dots show $1-\mathcal F^{'\text{m}}$ for the best controls obtained for each $T^\mathrm{m}$.
We find the upper bound for the quantum speed limit to be $T^{\mathrm{m}}_\text{QSL} \leq 0.0888 \,\si{ms}$
(see Appendix~\ref{app:oct} for the corresponding time-optimal control).
Compared to previous results of 0.97 fidelity \cite{de2008optimal,mundt2009optimalcontrol} (which also uses more approximations), this is still a factor ${\sim}1.69$ reduction in duration. Using instead 0.97 as the fidelity threshold the reduction is ${\sim}3.47$.
The blue translucent dots show $1-\mathcal F^{'\text{m}}$ for all optimized controls, where the translucency indicates the distribution density. From the translucency we see the average quality of the optimized controls increases with duration as the problem becomes easier (the seeming increase in low-fidelity solutions for the last three durations is due to an increased overall number of seeds). Nevertheless, the best obtained fidelities plateaus around 0.988 over a rather long interval.
As discussed previously, the full \sqswap gate can be realized by extending the merging optimized controls.
This is shown for the time-optimal controls at $T^{\mathrm{m}}_\text{QSL}$ in Fig.~\ref{fig:fullcontrolpropagation}  where instantaneous fidelities with various states are plotted as a function of time. We also plot the corresponding independently propagated single-particle wave functions \cite{Note1}. The initial state $\ket{\au_{Lg},\ad_{Rg}}$ is transferred into the merged trap in time $T^\mathrm{m}=0.0888\, \si{ms}$ with  $\mathcal F^{'\text{m}}=0.99$ and acquires a phase of $\alpha(T^\mathrm{m}) \approx 0.40$ during merging. The waiting stage with two-level dynamics lasts for $T\st{\sqswap} \approx 0.0375\, \si{ms}$
and nearly enters the entangled $\sqrt{\mathrm{SWAP}}\ket{\au_{e},\ad_{g}}$ state.
The controlled termination of the exchange interaction ensures that the remaining phase is acquired in the separation stage such that the final state is approximately $\sqrt{\mathrm{SWAP}}\ket{\au_{Lg},\ad_{Rg}}$ with fidelity $\mathcal{F^{\mathrm{f}}} = 0.983$. The total gate time is $T^{\mathrm{f}}=2T^{\mathrm{m}} + T\st{\sqswap} = 0.215\, \si{ms}$.
We note that the process is fundamentally limited by the rate at which entanglement can be generated. To provide a sense of scale,  $\ket{\au_e,\ad_g}$ prepared in the merged configuration would achieve $\alpha=0\rightarrow \pi/2$ in $T_\mathrm{\sqrt{SWAP}}=0.078\,\si{ms}$, which illustrates that there is still room for improvement.

The top row of table~\ref{tab:merit} summarizes several figures of merit for various states when propagated along the time-optimal control.
The superscripts m and f indicate if the quantity is measured at the merging or full gate duration, respectively.
Here,
$\mathcal F^{\text{m}}_{Lg\rightarrow e}$ and $\mathcal F^{\text{m}}_{Rg\rightarrow g}$ are the single-particle fidelities corresponding to the merging transfer $\ket{Lg}\ket{Rg} \rightarrow \ket{e}\ket{g}$,
$\mathcal F^{\text{f}}_{Lg\rightarrow Lg}$ and $\mathcal F^{\text{f}}_{Rg\rightarrow Rg}$ are the single-particle fidelities corresponding to the transfer $\ket{Lg}\ket{Rg} \rightarrow \ket{Lg}\ket{Rg}$ when extending the merging optimized control to the full gate,
$\mathcal{F^{\text{m}}_\pm}$ are the merging fidelities when including only the triplet ($+$) or singlet ($-$) component in the initial and target state, and
$\mathcal{F}^{\text{m}}_{\alpha_\text{t} = 0}$ is the merging fidelity if the target phase is excluded.

The high single-particle fidelities $\mathcal F^{\text{m}}_{Lg\rightarrow e}$ and $\mathcal F^{\text{m}}_{Rg\rightarrow g}$ indicate that the imperfection in $\mathcal F^{'\text{m}}$ is mainly due to the interaction affecting the triplet and singlet components differently.
A supporting observation is that $\mathcal F^{\text{m}}_- \gtrsim  0.99$ while $\mathcal F^{\text{m}}_+ \lesssim  0.99$.
This shows that relative phase acquired during the merging is a small but significant effect for producing high quality solutions.
Additionally, from comparing $\mathcal{F}^{\text{m}}_{\alpha_\text{t} = 0}$ and $\mathcal{F}^{\text{m}}$ we explicitly see that fidelity is not the best figure of merit for the merging sub-problem, but using it for optimization is still good enough when using $\mathcal{F}^{'\text{m}}$ as a stopping condition.

In Fig.~\ref{fig:FT_dechiara_full} we show optimization results for the full gate problem, using optimized merging controls as seeds.
2323 seeds were optimized while the remaining 221 were left out due to $\alpha(T^{\mathrm{m}}) \geq \pi/4$.
The red dots show $1-\mathcal F^\mathrm{f}$ for the monotonically best and optimal solutions, as there are now many unique durations (depending on  $\alpha(T^{\mathrm{m}})$). We find the upper bound for the quantum speed limit for the full gate to be $T^{\mathrm{f}}_\text{QSL} \leq 0.1377 \,\si{ms}$ (see Appendix~\ref{app:oct} for the corresponding time-optimal control).
This is faster than $T^{\mathrm{m}}=0.15\, \si{ms}$ for merging alone from Refs. \cite{de2008optimal,mundt2009optimalcontrol} and faster than $T^\mathrm{f}=0.215$ using the extended optimal merging solution.
Additionally, the sub-optimal plateau is now completely absent and we find an increase in optimal solutions by two orders of magnitude.
The blue dots show $1-\mathcal F^\mathrm{f}$ for all solutions.

\begin{figure*}[h]
	\includegraphics[scale=0.98]{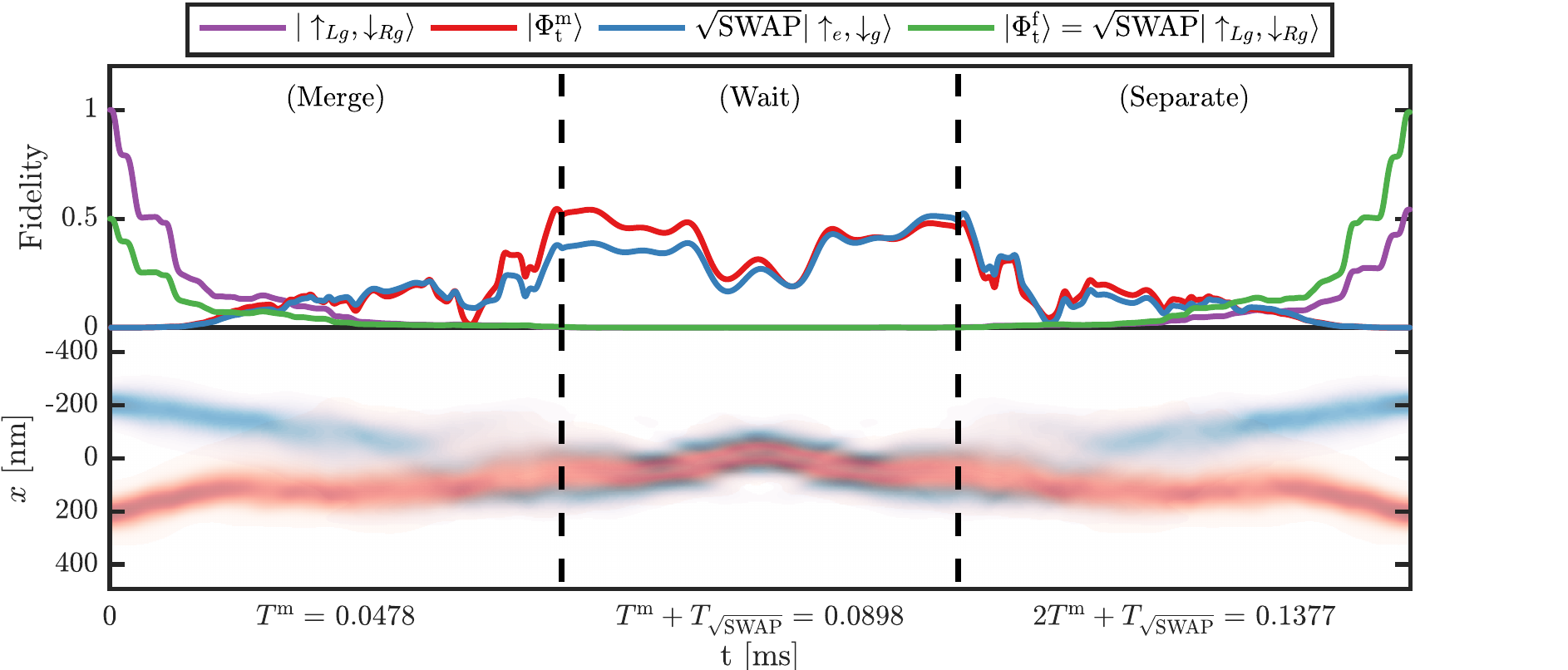}
	\caption{Full \sqswap gate operation from the time-optimal full gate control.
		The initial merge-wait-separate stages for the seed are indicated with the vertical dashed lines.
		Top: Instantaneous fidelities with various states. Bottom: Corresponding independent single-particle densities.
		The merging target state $\ket{\Phi^{\mathrm{m}}_\mathrm{t}}$ is only partially populated with at most $\mathcal F^{'\text{m}}\approx0.52$.  The atoms exhibit in-phase oscillations before the separation onset and enters  $\ket{\Phi^{\mathrm{f}}_\mathrm{t}}$ with $\mathcal F^{\text{f}}=0.99$.}
	\label{fig:fullcontrolpropagation_full}
\end{figure*}

Fig.~\ref{fig:fullcontrolpropagation_full} shows fidelities and independent single-particle densities when propagated along the time-optimal control at $T^{\mathrm{m}}_\text{QSL}$ \cite{Note1}.
Although there are no explicit distinct stages after optimizing, remnants of the seed's merge-wait-separate sequence (durations marked for reference) is still visible. For example, the time-optimal control as well as single-particle densities remain essentially symmetric around the $T^{\mathrm{f}}/2$.
The bottom row of table~\ref{tab:merit} summarizes figures of merit for various states.
The situation is quite different from Fig.~\ref{fig:fullcontrolpropagation} as the Hilbert space state trajectory is never close to passing through $\ket{\Phi^{\text{m}}_\text{t}}$ (or equivalently $\ket{e}\ket{g}$ in the independent-particle picture).
Nevertheless, the atomic overlap and interaction around $T^{\mathrm{f}}/2$ remains appreciable as the motional excitations allows for synchronized in-phase oscillations.
Finally, $\ket{\Phi^{\text{m}}_\text{t}}$ is prepared with $\mathcal F^\mathrm{f}=0.99$ at the end duration.
Interestingly, the seed that optimized to the time-optimal control initially had only $\mathcal F^{'\mathrm{m}} = 0.93$ at $T^\mathrm{m}=0.0478\, \si{ms}$. These numerical observations evidence that the restrictions imposed by the merge-wait-separate approach are indeed unnecessarily restrictive.

The plateau in Fig.~\ref{fig:FT_dechiara} suggests the existence of optimal merging controls at even lower durations, which could be uncovered by increasing the number of seeds and more elaborate seeding strategies.
However, the fact remains that such solutions can practically always be further improved by subsequently optimizing the full gate.
It would be more interesting to investigate the full gate with completely independently generated seeds as the current optimal solutions inherits at least partially the merge-wait-separate stages.
Additionally, the single-particle densities seem to have low velocities near the beginning and the end of the transfer. One can imagine rapidly transforming into the single-well configuration: both atoms are then subject to a large initial acceleration towards each other and will evolve without changing their shape appreciably as they closely resemble coherent states in a harmonic oscillator. As the atoms approach each other, suppose that the atoms could be decelerated to low momenta such that they oscillate out of phase with low amplitudes around the trap center. This would allow rapid entanglement since identical shapes maximize the interaction Eq.~\eqref{eq:Uab}. If the necessary relative phase is acquired within a little less than a single or few oscillation periods,
each atom will already have correctly directed opposite momenta at the onset of separation.
This is different from the case of in-phase oscillations in Fig.~\ref{fig:fullcontrolpropagation_full} where both atoms must be more delicately provided opposite accelerations to correctly separate them.
A similar scheme to the above is presented in Refs.~\cite{calarco2000quantum,treutlein2006microwave} for another type of collisional gate.
Whether such strategies are indeed effective in producing better solutions can only be verified by further numerical investigations and is left for future work.

\section{Conclusions}
\label{sec:conclusions}
We have presented and discussed in great detail the theory behind collisional \sqswap gate implementation in cold atoms. Prime among these considerations is proper accounting of the relative phase acquired during merging. Additionally, we argued that optimizing the full gate directly instead of the staged merge-wait-separate sequence is favorable both in terms of final fidelity and optimal trajectories in Hilbert space.
The relative phase and the full gate optimization allow separate reductions of the overall gate duration. Both concepts transcend any particular geometry, atomic species, and model dimensionality. They may thus be relevant in future work.

We then verified these claims in an optical lattice geometry. For the merging sub-problem, we find $\mathcal  F^{'\mathrm{m}}=0.99$ at $T^{\mathrm{m}}_\text{QSL} = 0.888\,\si{ms}$ which is an improvement over previous similar results of $\mathcal F^{'\mathrm{m}}=0.97$ at $T^{\mathrm{m}} = 0.15\,\si{ms}$. Nevertheless, a sub-optimal plateau of solutions indicate that the merging sub-problem is hard to solve and even in the best case the corresponding full gate fidelity is sub-optimal at $\mathcal F^{\mathrm{f}}=0.983$ in $T^{\mathrm{f}}=0.215\,\si{ms}$. By instead using the merging optimized solutions as seeds for the full gate problem the sub-optimal plateau is eliminated while also yielding a significantly increased number of optimal $\mathcal F^{\mathrm{f}}=0.99$ solutions with durations as low as $T^{\mathrm{f}}_\text{QSL} = 0.1377\,\si{ms}$.

\section{Acknowledgements}
This work was funded by the European Research Council, John Templeton Foundation, and the Carlsberg Foundation.
The numerical results presented in this work were obtained at the Centre for Scientific Computing, Aarhus http://phys.au.dk/forskning/cscaa/.

\clearpage

\makeatletter\onecolumngrid@push\makeatother
\begin{table*}[h]
		\begin{tabular*}{0.9\linewidth}{@{\extracolsep{\fill}} c | c c c c| c c c c c c | c c}
			&	& \multicolumn{2}{c}{\textbf{Single-particle}} & & & \multicolumn{4}{c}{\textbf{Two-particle}} && \\
			&
			$\mathcal F^{\text{m}}_{Lg\rightarrow e}$ &
			$\mathcal F^{\text{m}}_{Rg\rightarrow g}$ &
			$\mathcal F^{\text{f}}_{Lg\rightarrow Lg}$ &
			$\mathcal F^{\text{f}}_{Rg\rightarrow Rg}$ &
			$\mathcal F^{\text{m}}_{-} $ &
			$\mathcal F^{\text{m}}_{+}$ &
			$\mathcal{F}^{\text{m}}_{\alpha_\text{t} = 0}$ &
			$\mathcal{F}^{\text{m}}$ &
			$\mathcal F^{'\text{m}}$ &
			$\mathcal{F}^{\text{f}}$ &
			$T^{\mathrm{m}}$		 &
			$T^{\mathrm{f}}$
			\\
			\hline
			Merging optimized&  0.994 & 0.997  & 0.992 & 0.992 &0.992 & 0.988  & 0.950 & 0.9886 &0.990 & 0.983 & $0.0888\,\si{ms}$ & $0.215\,\si{ms}$\\
			Full gate optimized&  0.540 & 0.868 & 0.997 & 0.996 & 0.570 & 0.473  & 0.498 & 0.519 & 0.522 & 0.990  & $0.0478\,\si{ms}$ & $0.1377\,\si{ms}$\\
		\end{tabular*}
		\caption{Figures of merit when propagating different states along the same optimized controls corresponding either to the time-optimal merging control (top row) or time-optimal full gate control (bottom row).}
		\label{tab:merit}
\end{table*}

\clearpage
\makeatletter\onecolumngrid@pop\makeatother

\appendix

\section{Effective 1D Description}
\label{app:1ddesc}
Denoting by ($\vec r_i,s_i$) the spatial and spin degrees of freedom for the $i$'th particle, the full 3D Hamiltonian describing two interacting spin-$\frac{1}{2}$ particles is
\begin{align}
\op H\st{3D} =\op T +  U(\vec r_1) +  U(\vec r_2) +  U\st{int}(\vec r_1 - \vec r_2), \label{eq:3DHam}
\end{align}
where $\op T$ is the sum of kinetic energy operators over all particle coordinates, $ U(\vec r)$ is the single-particle trapping potential, and $ U\st{int} = g_{\text{3D}} \delta(\vec r_1 - \vec r_2)$ is the interaction potential.
Exactly solving the associated equations of motion for ${\Phi(\vec r_1,s_1,\vec r_2,s_2,t)}$  is computationally expensive even for very crude spatial discretizations.
For this reason, it is desirable to describe the approximate dynamics in an effective 1D model in which the spin states are treated implicitly.
This can be done under the assumption that motion in the remaining spatial axes are frozen out and the Hamiltonian is void of spin terms.
In this approximation, the variables separate into the product form
\begin{align}
 {\Psi^{(x)}(x_1,x_2,t)} {\Psi_{\text{gs}}^{(y)}(y_1,y_2)} {\Psi_\text{gs}^{(z)}(z_1,z_2)}{\chi(s_1,s_2)}, \label{eq:productform}
\end{align}
where ${\Psi^{(x)}(x_1,x_2,t)}$ is the only part of the wave function with a time evolution different from a trivial phase. The motional wave functions in the $y$ and $z$-directions remain in their respective ground state at all times, and the spin wave function  remains unchanged.
We can then restrict our attention to the non-trivial part of the state $\Phi \sim {\Psi^{(x)}}$ and drop the superscript as we have done throughout the main text.
We briefly discuss the steps to obtain \eqref{eq:productform} in the following.

The spin degrees of freedom separate exactly since there is no spin dependence in Eq.~\eqref{eq:3DHam}.
The spatial coordinates cannot be separated immediately because $U\st{int}$ couples them all. To proceed, we define $V(x) \equiv U(\vec r) \big|_{\vec r=\vec x}$ where $\vec x\equiv(x,0,0)$ and approximate the potential in $y$ and $z$ to be locally harmonic. This allows an approximate 1D description \cite{olshanii} of the inter-particle coupling
\begin{align}
g_\text{3D} = \frac{4a_s\pi \hbar}{m}&\rightarrow g_{\text{1D}} = 2a_s\hbar \sqrt{\omega_y\omega_z}, \label{app:app1}\\
\delta(\vec r_1-\vec r_2) &\rightarrow \delta(x_1-x_2).
\end{align}
Importantly, the local harmonic frequency $\omega_z$ ($\omega_y$) in the $\hat z$($\hat y$)-direction may become position dependent in $x$. To calculate these frequencies $U(\vec r)$ is Taylor expanded to second order around the point $\vec x$. Assuming that $\vec x$ is a minimum in $\hat y$ and $\hat z$ one obtains
\begin{align}
U(\vec r) &\approx U(\vec x) + \sum_{q=y,z} \bigg[ \frac{1}{2}\partial_{qq} U(\vec r)\cdot q^2\bigg]  \bigg|_{\vec r=\vec x} \nonumber
\\
&= V(x)  +  \sum_{q=y,z} \frac{1}{2}m\omega_q^2 q^2. \label{app:app3}
\end{align}
Comparing the two expressions we obtain the frequencies $\omega_q^2 = {\partial_{qq}U(\vec x)}/{m}$.
The full 3D potential for the optical lattice \cite{anderlini2006controlled,anderlini2007controlled,de2008optimal}
and corresponding frequencies are
\begin{widetext}
\begin{align}
U(\vec r) &= -V_z\cos^2k z  -V_0\bigg[\cos^2\left(\frac{\beta}{2}\right)\left\{\cos^2(ky)+\cos^2(k x-\frac{\pi}{2})\right\} + \sin^2\left(\frac{\beta}{2}\right)\left\{\cos(ky)+\cos(k x -\theta-\frac{\pi}{2})\right\}^2\bigg], \label{eq:3DU}\\
\omega_z & = \sqrt{\frac{2V_z k^2}{m}}, \qquad
\omega_y(x)  = \sqrt{\frac{2V_0 k^2}{m} \left[\cos^2\left(\frac{\beta}{2}\right) + \sin^2\left(\frac{\beta}{2}\right) \left\{ 1+ \cos(kx-\theta - \frac{\pi}{2}) \right\}\right]}. \label{eq:localharmfreq}
\end{align}
\end{widetext}
As $\omega_z$ does not depend on $x$, the $z$-degrees of freedom separate exactly in these approximations. Excitations along this axis can always be suppressed by choosing
the independent trap depth $V_z$ to generate sufficiently large vibrational frequencies with associated energy spacings.
On the contrary, the separation of the $x$ and $y$ coordinates is only approximate since $\omega_y=\omega_y(x)$.
Our calculations of the full 2D single-particle spectrum show that the trapping along $x$ is only slightly weaker than along $y$, since $V_0$ is common to both axes. It is therefore much harder to suppress excitations along $y$ than $z$.
Errors induced by the approximate potential separability of $x$ and $y$ is the main limitation of the model, since this coupling is much larger than the inter-particle coupling $U\st{int}$.
The quality of the approximation \eqref{eq:productform} can thus be assessed on the independent-particle level.
In Fig.~\ref{fig:2dcompare} we propagate the particle starting in the left ground state over the time-optimal control at $T^{\mathrm{f}}_{\text{QSL}}$ and compute the 1D and 2D instantaneous fidelities with the initial state. Their difference corresponds roughly to the leakage out of the ground state in the $y$-direction induced by the non-separability of the potential. The effects are less pronounced for the particle starting in the right ground state. We have done the same for the optimized control presented in Fig.~7 from Ref. \cite{de2008optimal} and find similar results.
\begin{figure}[t]

\end{figure}

\section{Methods} \label{app:methods}

\subsection{Numerics} \label{app:numerics}
The simulations and optimization results presented in this paper are produced with the QEngine \cite{sorensen2019qengine}, our recent C++ software package for quantum optimal control.

The singlet and triplet states $\ket{\Phi^\pm}\sim \ket{\Psi^\pm} $ are obtained through numerical diagonalization in the discrete $\ket{x_1,x_2}$-representation with uniform grid-spacing. The 5-diagonal approximation is used for the Laplacian.
Propagation is performed using split-step FFT, which is fastest if the number of spatial grid points are integer powers of 2, e.g. $D \in \{32,64,128,256,512,1024\}$.
We employ a state absorptive imaginary potential near the grid borders to minimize the effects of the periodic boundary conditions induced by the split-step FFT method.
The $\delta(x_1-x_2)$ in the interaction potential necessitates a very high degree of temporal resolution to reliably produce the correct dynamics.
Numerical experiments show that $\delta t = 1.2\cdot 10^{-5}$ is a good value for the simulation. At this very high temporal resolution the spatial resolution is found to be stable over a surprisingly broad range. Propagating thousands of different controls, the mean and standard deviation measured between fidelities produced with $D\st{max}=1024$ and $D=64$ is roughly $(2.2\pm 2.7)\cdot 10^{-4}$. Increasing $D$ has almost no effect. However, even at this relatively modest $D$ the time discretization still requires $\mathcal{O}(10^4)$ steps to reach the durations of interest and as a result performing optimization with these parameters is very slow. To significantly speed up the optimization by several orders of magnitude we instead use increasingly finer grids. A grid is defined by the tuple $\{D,\delta t\}$.
For the merging sub-problem, we optimize on the grids sequentially $\{32,5 \cdot 10^{-4}\} \rightarrow \{64, 1 \cdot 10^{-4}\} \rightarrow \{64, 1.2 \cdot 10^{-5}\}$.
The control is interpolated linearly to the new $\delta t$ when moving between the grids. This allows performing a large scale multistarting optimization at a broad range of $T$ on the approximate time scale of several days.
Due to memory issues we only optimize the full gate on the grids $\{32,5 \cdot 10^{-4}\} \rightarrow \{64, 1 \cdot 10^{-4}\}$ and then simply evaluate on the final grid $\{64, 1.2 \cdot 10^{-5}\}$ when reporting results.

\subsection{Optimal Control} \label{app:oct}
To solve the state transfer problem $\ket{\Phi_0} \rightarrow \ket{\Phi_\text{t}}$, we use the $L^2$ gradient based \textsc{grape} algorithm with the \textsc{l-bfgs} search direction to minimize the cost functional
\begin{align*}
J[\mathcal U] &= J_\mathcal{F}+ J_\gamma + J_\sigma \\
&=\frac{1}{2}\left(1-\mathcal F\right) + \sum_{i=1}^{k}\left[\frac{\gamma}{2}\int_0^T\dot u_i^2 \d t  + \frac{\sigma}{2}\int_0^T b(u_i) \d t \right]
\end{align*}
by iteratively improving the set of $k$ control fields (protocols) $\mathcal U(t) = \{u_i(t)\}_{i=1}^k$. The gradient is calculated using the adjoint method by introducing an additional Lagrange multiplier term.
Minimizing $J_\mathcal{F}$ corresponds to maximizing the fidelity $\mathcal F = |\braket{\Phi_\text{t} | \Phi(T)}|^2 = |\braket{\Phi_\text{t} | \hat U(\mathcal{U}) | \Phi_0}|^2$ where $\hat U$ is the time evolution operator.
 $J_\gamma$ adds preference  to smoother controls with strength $\gamma$ and $J_\sigma$ adds preference to controls within specified parameter boundaries $u_i(t) \in [u\st{min},u\st{max}]_i$ with strength $\sigma$. In the latter context, $b(u)$ is a function that is zero when the boundaries are respected and parabolic when exceeded \cite{sorensen2019qengine}. \\

As mentioned in the main text, a more appropriate measure of the transfer quality for the merging sub-problem is the total population
$\mathcal{F}' = |\braket{\tilde \Phi^+_{e,g}| \Phi(T)}|^2 + |\braket{\tilde \Phi^-_{e,g}| \Phi(T)}|^2$.
This could be directly accommodated in the cost functional by replacing $J_\mathcal{F}$ with
\begin{align*}
\frac{1}{2}\left(1-\mathcal F' + \frac{1}{2}\left\{ \frac{|\braket{\tilde \Phi^+_{e,g}| \Phi(T)}|^2}{2} - \frac{|\braket{\tilde \Phi^-_{e,g}| \Phi(T)}|^2}{2}\right\}^2\right)
\end{align*}
which is minimized when $\ket{\Phi(T)}$ is fully and equally distributed in the states $\ket{\tilde \Phi^\pm_{e,g}}$ independent of relative phase, see Eq.~\eqref{eq:mtarget}. This replacement requires calculating new G\^ ateaux derivatives to obtain a new optimality system \cite{sorensen2019qengine}.
Instead of formally doing this we simply use $\mathcal F'$ as a stopping condition. \\

We simulate rubidium atoms with mass $m_{Rb} = 87\,\si{amu}$ and assume a state independent scattering length $a_{Rb} = 5.45\,\si{nm} \approx 103 a_0$ \cite{kaufman2014two,julienne1997collisional} where $a_0$ is the Bohr radius.
We restrict our attention to a single unit cell of the lattice defined within $x\in  [-1.0, +1.0]\times a$ where $a=408=\lambda/2$ \cite{anderlini2007controlled} is the lattice site separation for the initial configuration Fig.~\ref{fig:VLattice}\textbf{(a)}. We pad the boundaries slightly such that $x\in  [-1.2, +1.2] \times a$ for numerical reasons: in the boundary region a constant $V\st{cst} = \max \,V(x)$ is used to stabilize the diagonalization.
We have verified that the wave function does not enter this non-physical region when propagated along the optimized controls.
The independent lattice in the $z$-direction has strength  $V_z = 186\, \si{kHz}\cdot h$ \cite{anderlini2007controlled}.


The control parameters of the potential Eq.~\eqref{eq:VLattice} are rescaled as
$\mathcal U = \{\beta(t),\theta(t),V_0(t)\} \rightarrow \{\beta(t), \theta(t), V_0(t)\} \times  \{\beta\st{scale},   \theta\st{scale}, V_{0,\text{scale}}\}$ with values
\begin{align*}
\{\beta\st{scale},   \theta\st{scale}, V_{0,\text{scale}}\} &= \{0.52\pi,-0.474\pi, 122\si{kHz}\cdot h\} \\
\mathcal U(0) &=\{\beta(0), \; \theta(0), \, V_0(0)\}\, =\{0, 1, 1 \} \\
\mathcal U(T) &=\{\beta(T), \theta(T), V_0(T)\}=\{1, 1, 1 \}
\end{align*}
The unscaled initial and final control values for the merging sub-problem ($\beta=0\rightarrow 0.52\pi$, $\theta=-0.474\pi$, $V_0=122\,\si{kHz}\cdot h$) are chosen to allow comparison with results in Ref. \cite{de2008optimal}.
Similar values are used in Ref. \cite{mundt2009optimalcontrol}.
Neither paper discuss rescaling.
For the full gate problem $\mathcal U(T) = \mathcal U(0)$.

\begin{figure*}[t]
	\includegraphics[]{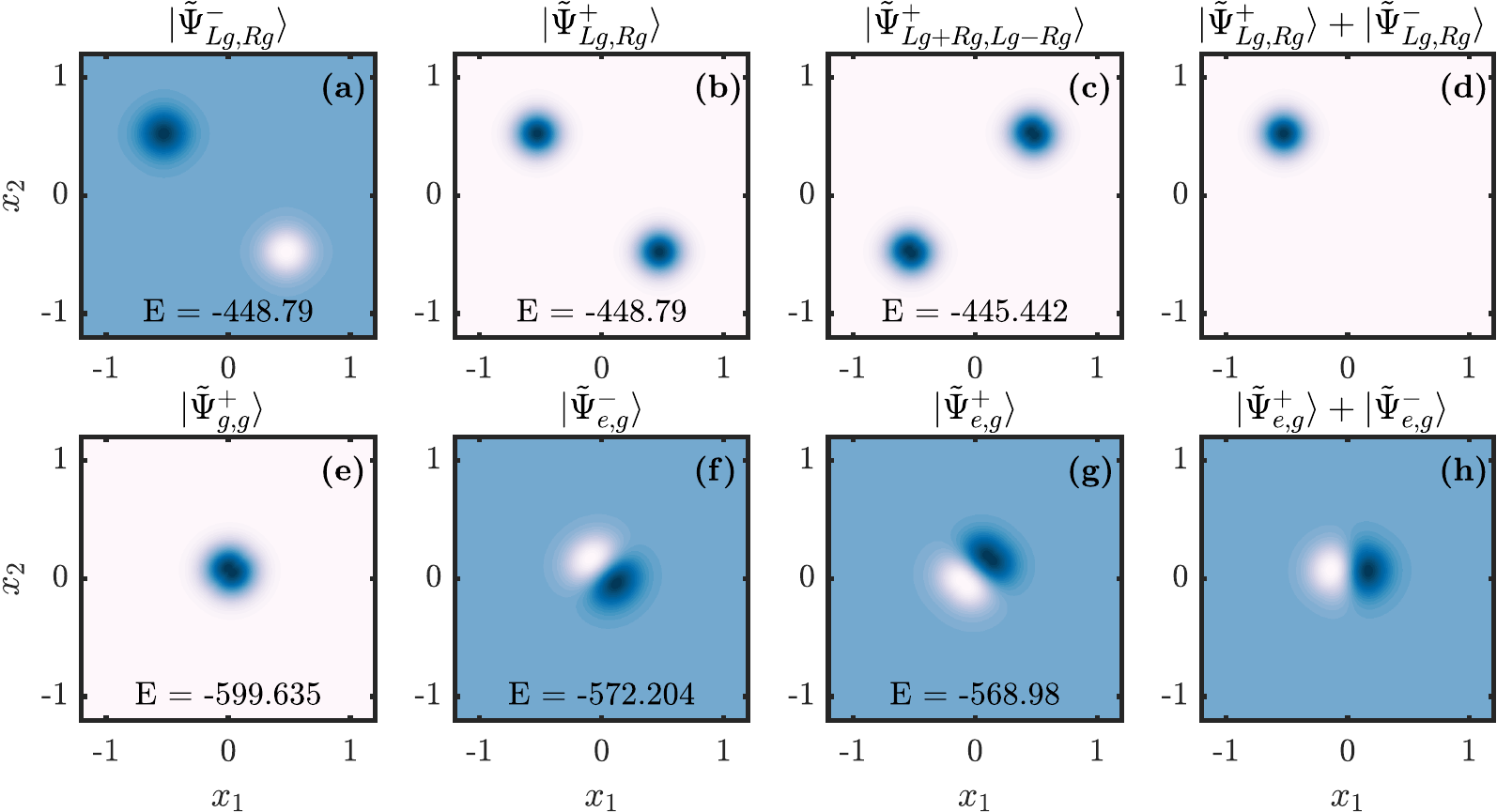}
	\caption{Numerical two-particle states in initial configuration (top row) and merged configuration (bottom row). Energies are in units of $\si{kHz}\cdot h$. \textbf{(a)}-\textbf{(c)} Lowest lying eigenstates. The (ground) states $\ket{\tilde \Psi^\pm_{Lg,Lg}}$ corresponding to singly occupied wells are degenerate since both wave functions are vanishing for $x_1=x_2$. The excited state $\ket{\tilde \Psi^\pm_{Lg+Rg,Lg-Rg}}$ corresponding to doubly occupied wells has an increased energy due to the interaction. \textbf{(d)} Numerical initial state $\ket{ \Psi_{0}}$.
		\textbf{(e)}-\textbf{(g)} Lowest lying eigenstates. The symmetric excited state $\ket{\tilde \Psi^+_{e,g}}$ has an increased energy   compared to the antisymmetric $\ket{\tilde \Psi^-_{e,g}}$. \textbf{(h)} Numerical target state $\ket{ \Psi^{\text{m}}_{\text{t}}}$ (here $\alpha\st{t}=0$).
}
\label{fig:dechiara_spectrum}
\end{figure*}

Fig.~\ref{fig:dechiara_spectrum} shows a few select two-particle spatial states.
For the separated configuration Fig.~\ref{fig:VLattice}\textbf{(a)}, $\ket{\tilde \Psi^\pm_{Lg,Lg}}$ (singly occupied wells) are degenerate, while the interaction increases the energy of $\ket{\tilde \Psi^+_{Lg+Rg,Lg-Rg}}$ (doubly occupied wells) by ${\sim}3\,\si{kHz}\cdot h$ as was also reported in Ref. \cite{de2008optimal}. Note that the numerical initial state $\ket{ \Psi_{0}}$ and target state $\ket{ \Psi^\mathrm{m}_{\text{t}}}$ are not immediately symmetric, but are still physically correct since the appropriate spin degree of freedom is implicitly included.

\begin{figure*}[t]
	\begin{minipage}[t]{0.483\linewidth}
		\includegraphics[]{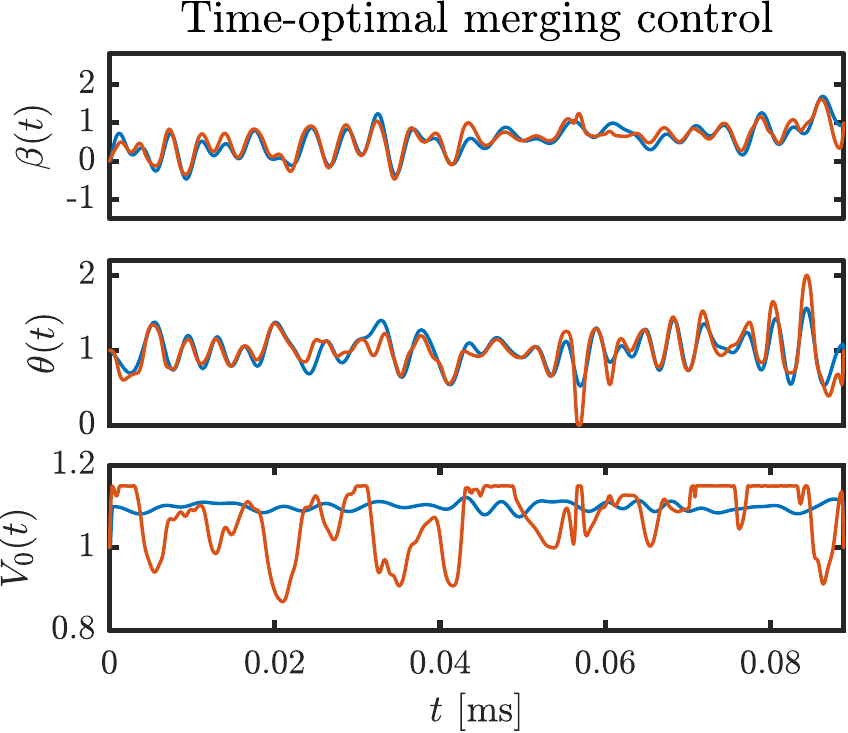}
		\caption{(Color online) The set of (scaled) optimal controls 
			and their seed at the quantum speed limit bound $T^{\mathrm{m}}_\text{QSL} = 0.0888 \,\si{ms}$. Blue: Initial control. Red: Optimized control.}
		\label{fig:control_dechira}
	\end{minipage}
	\hspace{0.4cm}
	\begin{minipage}[t]{0.483\linewidth}
		\includegraphics[]{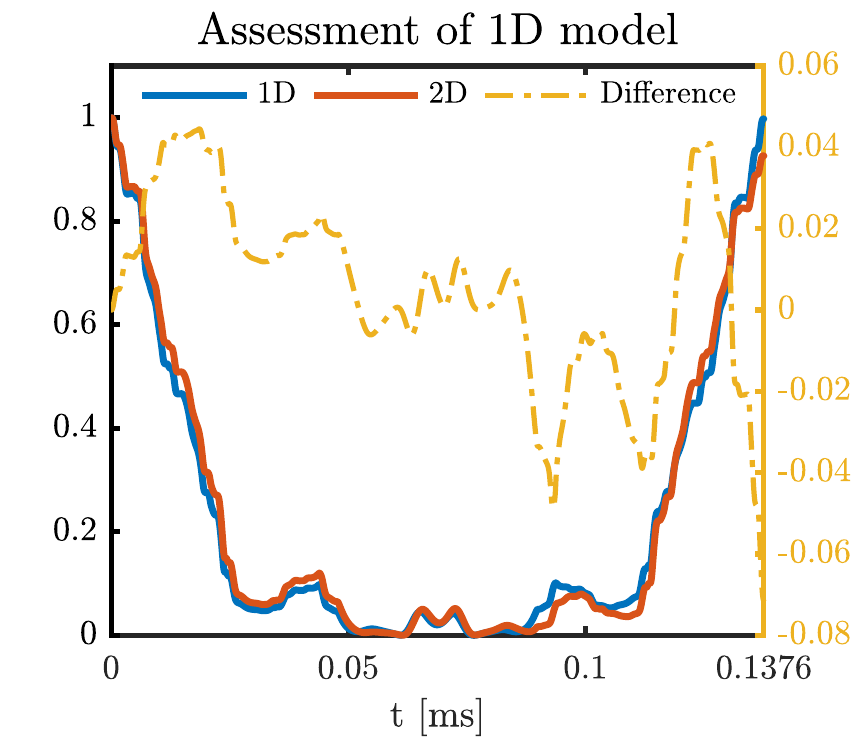}
		\caption{(Color online) Comparison of instantaneous fidelities with initial state in the 1D and 2D case. Their difference roughly corresponds to the leakage out of the ground state in the $\hat y$-direction.
		}
		\label{fig:2dcompare}
	\end{minipage}
\end{figure*}

\begin{figure*}[h]
	\includegraphics[]{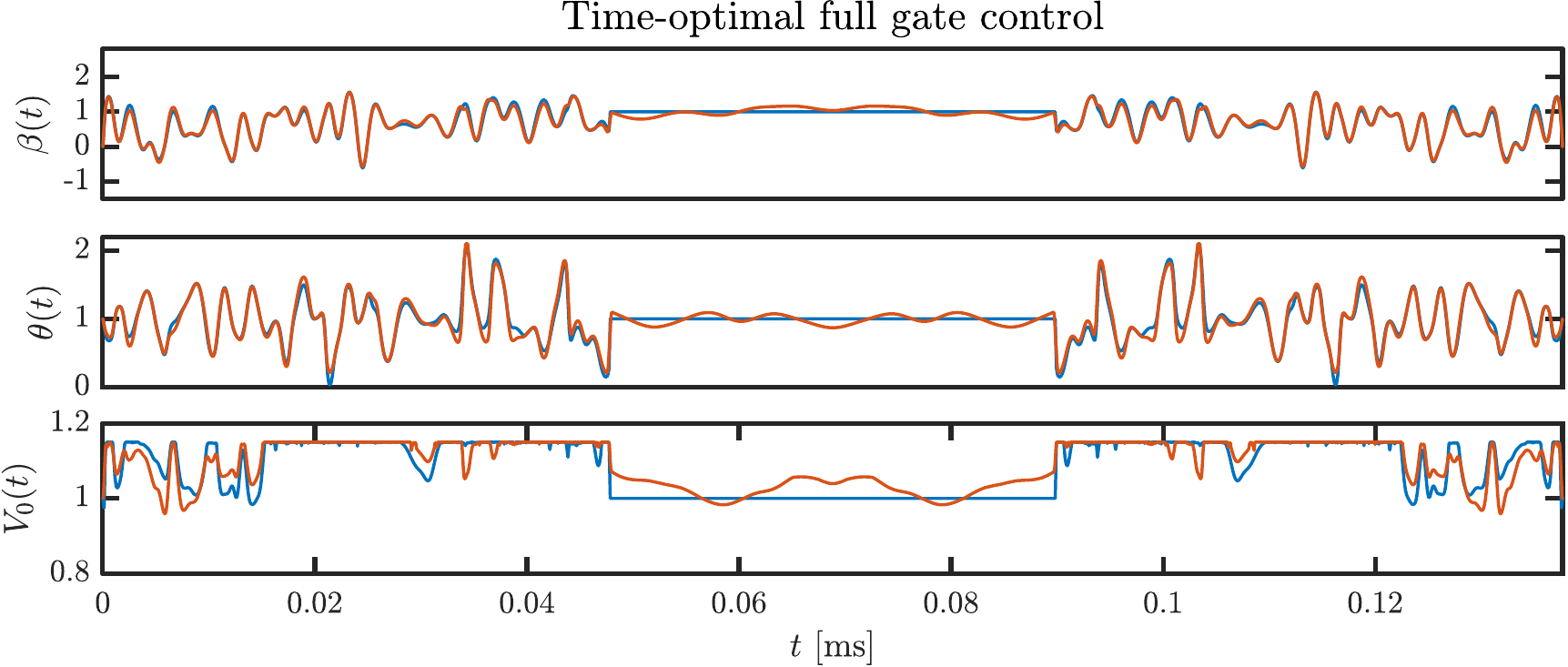}
	\caption{(Color online) The set of (scaled) optimal controls
		and their seed at the quantum speed limit bound $T^{\mathrm{f}}_\text{QSL} = 0.1377 \,\si{ms}$. Blue: Initial control. Red: Optimized control.
	}
	\label{fig:control_dechira_full}
\end{figure*}

We include a regularization term with $\gamma=10^{-7}$ for all controls and a boundary term to constrain
$\{-\infty, 0, 0.2\} \leq \{\beta(t), \theta(t), V_0(t)\} \leq \{+\infty, 2.1, 1.15\}$ with $\sigma=10^5$ such that the adjacent lattice unit cells do not mix and $V_0$ remains reasonably lower than $V_z$ (see Appendix \ref{app:1ddesc}).
Optimization seeds for merging are generated by perturbing a reference control with $M \sim 40-60$ random sines of increasing harmonic frequency with random weighting and overall normalization. The reference control was heuristically chosen such that the seed cost is on average decreased.
As discussed in the main text, the optimized merging controls are used as seeds for the full gate optimization. In both cases we optimize the seeds until they exceed the figure of merit threshold, converge to a local minimum, or exceed a wall time limit of 7 days.
Figs.~\ref{fig:control_dechira} and \ref{fig:control_dechira_full} shows the time-optimal controls for the merging and full gate problems, respectively \cite{Note1}.

\makeatletter\onecolumngrid@push\makeatother
\clearpage
\makeatletter\onecolumngrid@pop\makeatother

\makeatletter\onecolumngrid@push\makeatother

\clearpage
\makeatletter\onecolumngrid@pop\makeatother

\section{Spin Exchange Hamiltonian}
\label{appendix:spinham}
The collisional effects modeled by Eq. \eqref{eq:H} is purely spatial which allowed us to treat the spin states implicitly. However, if the trapping geometry is static and the spatial state is a superposition on the form $\ket{\tilde \Phi^+_{a,b}} + \ket{\tilde \Phi^-_{a,b}}$ we can describe the dynamics with an effective spin Hamiltonian $\op H_{\text{spin}}= J_{\text{ex}}\cdot \op{\vec{S}}_1 \otimes \op {\vec{S}}_2$ where $\op{\vec {S}}_i$ are spin operators and $J_{\text{ex}}$ is the exchange energy.
The matrix representation of  $\op H_{\text{spin}}$ in the standard computational basis $\left\{ \ket{\ad,\ad} \doteq \vec e_1, \ket{\au,\ad} \doteq \vec e_2,\ket{\ad,\au} \doteq \vec e_3 ,\ket{\au,\au} \doteq \vec e_4\right\}$ is
\begin{align}
\hat H_{\text{spin}} & \doteq \frac{J_{\text{ex}}\hbar^2}{4}
\begin{bmatrix}
1 & 0 & 0 & 0\\
0 & -1 & 2 & 0\\
0 & 2 & -1 & 0\\	
0 & 0 & 0 & 1	
\end{bmatrix},
\label{eq:spinHam}
\end{align}
where $\vec e_i$ are standard unit vectors.
Clearly the computational basis states are not eigenstates of $\op H_{\text{spin}}$ unless $J_{\text{ex}}=0$. Diagonalizing of Eq.~\eqref{eq:spinHam} yields energies and corresponding states
\begin{align}
&E^- = -\frac{3}{4}J_{\text{ex}}\hbar^2:  &&\vec \chi^- = \frac{1}{\sqrt{2}}(\vec e_2 - \vec e_3), \nonumber\\
&E^+ =+\frac{1}{4}J_{\text{ex}}\hbar^2:  &\vec e_1,\qquad&\vec \chi^+ = \frac{1}{\sqrt{2}}(\vec e_2 + \vec e_3),\qquad \vec e_4, \nonumber
\end{align}
with three degenerate states corresponding to $E^+$.
The $\vec \chi^\pm$ states are exactly the spin singlet and triplet spin states with an energy difference of $\tilde U  =  J_{ex}$, while $\vec \chi^{\au\au} = \vec e_1$ and $\vec \chi^{\ad\ad} = \vec e_4$ are the remaining triplet spin states with a non-zero net spin. In this effective spin model, it is the spatial degree of freedom that is treated implicitly. Thus, the time evolution for the initially prepared state $\ket{\au_a,\ad_b} = \ket{\tilde \Phi^+_{a,b}} + \ket{\tilde \Phi^-_{a,b}}$ is
\begin{align}
\ket{\tilde \Phi(t)} &= e^{-\frac{i\op H_{\text{spin}}t}{\hbar}}\ket{\au_a,\ad_b} = e^{-\frac{i \op H_{\text{spin}}t}{\hbar}}\left[ \ket{\tilde \Phi^+_{a,b}} + \ket{\tilde \Phi^-_{a,b}}\right] \nonumber\\
&= \ket{\tilde\Psi^+_{a,b}}e^{-\frac{i\op H_{\text{spin}}t}{\hbar}}\ket{\chi^+} + \ket{\tilde\Psi^-_{a,b}}e^{-\frac{i\op H_{\text{spin}}t}{\hbar}}\ket{\chi^-} \nonumber\\
&= e^{-\frac{iE^+t}{\hbar}}\ket{\tilde\Psi^+_{a,b}}\ket{\chi^+} + e^{-\frac{iE^-t}{\hbar}}\ket{\tilde\Psi^-_{a,b}}\ket{\chi^-}\nonumber\\
&\rightarrow \ket{\tilde\Phi^+_{a,b}} +  e^{i\alpha(t)}\ket{\tilde \Phi^-_{a,b}} \nonumber
 \end{align}
where $\alpha(t) = J_{\text{ex}}t/\hbar$ and we ignored a global phase. This model exactly reproduces the dynamics from Eq. \eqref{eq:swap}.

	\bibliographystyle{apsrev_custom_nourl}
	\bibliography{references}

\begin{thebibliography}{50}
\expandafter\ifx\csname natexlab\endcsname\relax\def\natexlab#1{#1}\fi
\expandafter\ifx\csname bibnamefont\endcsname\relax
  \def\bibnamefont#1{#1}\fi
\expandafter\ifx\csname bibfnamefont\endcsname\relax
  \def\bibfnamefont#1{#1}\fi
\expandafter\ifx\csname citenamefont\endcsname\relax
  \def\citenamefont#1{#1}\fi
\expandafter\ifx\csname url\endcsname\relax
  \def\url#1{\texttt{#1}}\fi
\expandafter\ifx\csname urlprefix\endcsname\relax\def\urlprefix{URL }\fi
\providecommand{\bibinfo}[2]{#2}
\providecommand{\eprint}[2][]{\url{#2}}

\bibitem[{\citenamefont{Chu}(2002)}]{chu2002cold}
\bibinfo{author}{\bibfnamefont{S.}~\bibnamefont{Chu}},
  \bibinfo{journal}{Nature} \textbf{\bibinfo{volume}{416}},
  \bibinfo{pages}{206} (\bibinfo{year}{2002}).

\bibitem[{\citenamefont{Bloch et~al.}(2008)\citenamefont{Bloch, Dalibard, and
  Zwerger}}]{bloch2008many}
\bibinfo{author}{\bibfnamefont{I.}~\bibnamefont{Bloch}},
  \bibinfo{author}{\bibfnamefont{J.}~\bibnamefont{Dalibard}}, \bibnamefont{and}
  \bibinfo{author}{\bibfnamefont{W.}~\bibnamefont{Zwerger}},
  \bibinfo{journal}{Reviews of modern physics} \textbf{\bibinfo{volume}{80}},
  \bibinfo{pages}{885} (\bibinfo{year}{2008}).

\bibitem[{\citenamefont{Sherson et~al.}(2010)\citenamefont{Sherson, Weitenberg,
  Endres, Cheneau, Bloch, and Kuhr}}]{sherson2010single}
\bibinfo{author}{\bibfnamefont{J.~F.} \bibnamefont{Sherson}},
  \bibinfo{author}{\bibfnamefont{C.}~\bibnamefont{Weitenberg}},
  \bibinfo{author}{\bibfnamefont{M.}~\bibnamefont{Endres}},
  \bibinfo{author}{\bibfnamefont{M.}~\bibnamefont{Cheneau}},
  \bibinfo{author}{\bibfnamefont{I.}~\bibnamefont{Bloch}}, \bibnamefont{and}
  \bibinfo{author}{\bibfnamefont{S.}~\bibnamefont{Kuhr}},
  \bibinfo{journal}{Nature} \textbf{\bibinfo{volume}{467}}, \bibinfo{pages}{68}
  (\bibinfo{year}{2010}).

\bibitem[{\citenamefont{Weitenberg
  et~al.}(2011{\natexlab{a}})\citenamefont{Weitenberg, Endres, Sherson,
  Cheneau, Schau{\ss}, Fukuhara, Bloch, and Kuhr}}]{weitenberg2011single}
\bibinfo{author}{\bibfnamefont{C.}~\bibnamefont{Weitenberg}},
  \bibinfo{author}{\bibfnamefont{M.}~\bibnamefont{Endres}},
  \bibinfo{author}{\bibfnamefont{J.~F.} \bibnamefont{Sherson}},
  \bibinfo{author}{\bibfnamefont{M.}~\bibnamefont{Cheneau}},
  \bibinfo{author}{\bibfnamefont{P.}~\bibnamefont{Schau{\ss}}},
  \bibinfo{author}{\bibfnamefont{T.}~\bibnamefont{Fukuhara}},
  \bibinfo{author}{\bibfnamefont{I.}~\bibnamefont{Bloch}}, \bibnamefont{and}
  \bibinfo{author}{\bibfnamefont{S.}~\bibnamefont{Kuhr}},
  \bibinfo{journal}{Nature} \textbf{\bibinfo{volume}{471}},
  \bibinfo{pages}{319} (\bibinfo{year}{2011}{\natexlab{a}}).

\bibitem[{\citenamefont{Kaufman et~al.}(2014)\citenamefont{Kaufman, Lester,
  Reynolds, Wall, Foss-Feig, Hazzard, Rey, and Regal}}]{kaufman2014two}
\bibinfo{author}{\bibfnamefont{A.}~\bibnamefont{Kaufman}},
  \bibinfo{author}{\bibfnamefont{B.}~\bibnamefont{Lester}},
  \bibinfo{author}{\bibfnamefont{C.}~\bibnamefont{Reynolds}},
  \bibinfo{author}{\bibfnamefont{M.}~\bibnamefont{Wall}},
  \bibinfo{author}{\bibfnamefont{M.}~\bibnamefont{Foss-Feig}},
  \bibinfo{author}{\bibfnamefont{K.}~\bibnamefont{Hazzard}},
  \bibinfo{author}{\bibfnamefont{A.}~\bibnamefont{Rey}}, \bibnamefont{and}
  \bibinfo{author}{\bibfnamefont{C.}~\bibnamefont{Regal}},
  \bibinfo{journal}{Science} \textbf{\bibinfo{volume}{345}},
  \bibinfo{pages}{306} (\bibinfo{year}{2014}).

\bibitem[{\citenamefont{Wang et~al.}(2015)\citenamefont{Wang, Zhang,
  Corcovilos, Kumar, and Weiss}}]{wang2015coherent}
\bibinfo{author}{\bibfnamefont{Y.}~\bibnamefont{Wang}},
  \bibinfo{author}{\bibfnamefont{X.}~\bibnamefont{Zhang}},
  \bibinfo{author}{\bibfnamefont{T.~A.} \bibnamefont{Corcovilos}},
  \bibinfo{author}{\bibfnamefont{A.}~\bibnamefont{Kumar}}, \bibnamefont{and}
  \bibinfo{author}{\bibfnamefont{D.~S.} \bibnamefont{Weiss}},
  \bibinfo{journal}{Physical review letters} \textbf{\bibinfo{volume}{115}},
  \bibinfo{pages}{043003} (\bibinfo{year}{2015}).

\bibitem[{\citenamefont{Kim et~al.}(2016)\citenamefont{Kim, Lee, Lee, Jo, Song,
  and Ahn}}]{kim2016situ}
\bibinfo{author}{\bibfnamefont{H.}~\bibnamefont{Kim}},
  \bibinfo{author}{\bibfnamefont{W.}~\bibnamefont{Lee}},
  \bibinfo{author}{\bibfnamefont{H.-g.} \bibnamefont{Lee}},
  \bibinfo{author}{\bibfnamefont{H.}~\bibnamefont{Jo}},
  \bibinfo{author}{\bibfnamefont{Y.}~\bibnamefont{Song}}, \bibnamefont{and}
  \bibinfo{author}{\bibfnamefont{J.}~\bibnamefont{Ahn}},
  \bibinfo{journal}{Nature communications} \textbf{\bibinfo{volume}{7}},
  \bibinfo{pages}{13317} (\bibinfo{year}{2016}).

\bibitem[{\citenamefont{Endres et~al.}(2016)\citenamefont{Endres, Bernien,
  Keesling, Levine, Anschuetz, Krajenbrink, Senko, Vuletic, Greiner, and
  Lukin}}]{endres2016atom}
\bibinfo{author}{\bibfnamefont{M.}~\bibnamefont{Endres}},
  \bibinfo{author}{\bibfnamefont{H.}~\bibnamefont{Bernien}},
  \bibinfo{author}{\bibfnamefont{A.}~\bibnamefont{Keesling}},
  \bibinfo{author}{\bibfnamefont{H.}~\bibnamefont{Levine}},
  \bibinfo{author}{\bibfnamefont{E.~R.} \bibnamefont{Anschuetz}},
  \bibinfo{author}{\bibfnamefont{A.}~\bibnamefont{Krajenbrink}},
  \bibinfo{author}{\bibfnamefont{C.}~\bibnamefont{Senko}},
  \bibinfo{author}{\bibfnamefont{V.}~\bibnamefont{Vuletic}},
  \bibinfo{author}{\bibfnamefont{M.}~\bibnamefont{Greiner}}, \bibnamefont{and}
  \bibinfo{author}{\bibfnamefont{M.~D.} \bibnamefont{Lukin}},
  \bibinfo{journal}{Science} \textbf{\bibinfo{volume}{354}},
  \bibinfo{pages}{1024} (\bibinfo{year}{2016}).

\bibitem[{\citenamefont{Barredo et~al.}(2016)\citenamefont{Barredo,
  De~L{\'e}s{\'e}leuc, Lienhard, Lahaye, and Browaeys}}]{barredo2016atom}
\bibinfo{author}{\bibfnamefont{D.}~\bibnamefont{Barredo}},
  \bibinfo{author}{\bibfnamefont{S.}~\bibnamefont{De~L{\'e}s{\'e}leuc}},
  \bibinfo{author}{\bibfnamefont{V.}~\bibnamefont{Lienhard}},
  \bibinfo{author}{\bibfnamefont{T.}~\bibnamefont{Lahaye}}, \bibnamefont{and}
  \bibinfo{author}{\bibfnamefont{A.}~\bibnamefont{Browaeys}},
  \bibinfo{journal}{Science} \textbf{\bibinfo{volume}{354}},
  \bibinfo{pages}{1021} (\bibinfo{year}{2016}).

\bibitem[{\citenamefont{Lee et~al.}(2016)\citenamefont{Lee, Kim, and
  Ahn}}]{lee2016three}
\bibinfo{author}{\bibfnamefont{W.}~\bibnamefont{Lee}},
  \bibinfo{author}{\bibfnamefont{H.}~\bibnamefont{Kim}}, \bibnamefont{and}
  \bibinfo{author}{\bibfnamefont{J.}~\bibnamefont{Ahn}},
  \bibinfo{journal}{Optics express} \textbf{\bibinfo{volume}{24}},
  \bibinfo{pages}{9816} (\bibinfo{year}{2016}).

\bibitem[{\citenamefont{Kumar et~al.}(2018)\citenamefont{Kumar, Wu, Giraldo,
  and Weiss}}]{kumar2018sorting}
\bibinfo{author}{\bibfnamefont{A.}~\bibnamefont{Kumar}},
  \bibinfo{author}{\bibfnamefont{T.-Y.} \bibnamefont{Wu}},
  \bibinfo{author}{\bibfnamefont{F.}~\bibnamefont{Giraldo}}, \bibnamefont{and}
  \bibinfo{author}{\bibfnamefont{D.~S.} \bibnamefont{Weiss}},
  \bibinfo{journal}{Nature} \textbf{\bibinfo{volume}{561}}, \bibinfo{pages}{83}
  (\bibinfo{year}{2018}).

\bibitem[{\citenamefont{Barredo et~al.}(2018)\citenamefont{Barredo, Lienhard,
  de~L{\'e}s{\'e}leuc, Lahaye, and Browaeys}}]{barredo2018_3dassembly}
\bibinfo{author}{\bibfnamefont{D.}~\bibnamefont{Barredo}},
  \bibinfo{author}{\bibfnamefont{V.}~\bibnamefont{Lienhard}},
  \bibinfo{author}{\bibfnamefont{S.}~\bibnamefont{de~L{\'e}s{\'e}leuc}},
  \bibinfo{author}{\bibfnamefont{T.}~\bibnamefont{Lahaye}}, \bibnamefont{and}
  \bibinfo{author}{\bibfnamefont{A.}~\bibnamefont{Browaeys}},
  \bibinfo{journal}{Nature} \textbf{\bibinfo{volume}{561}}, \bibinfo{pages}{79}
  (\bibinfo{year}{2018}).

\bibitem[{\citenamefont{Saskin et~al.}(2019)\citenamefont{Saskin, Wilson,
  Grinkemeyer, and Thompson}}]{saskin2019narrow}
\bibinfo{author}{\bibfnamefont{S.}~\bibnamefont{Saskin}},
  \bibinfo{author}{\bibfnamefont{J.}~\bibnamefont{Wilson}},
  \bibinfo{author}{\bibfnamefont{B.}~\bibnamefont{Grinkemeyer}},
  \bibnamefont{and} \bibinfo{author}{\bibfnamefont{J.~D.}
  \bibnamefont{Thompson}}, \bibinfo{journal}{Physical review letters}
  \textbf{\bibinfo{volume}{122}}, \bibinfo{pages}{143002}
  (\bibinfo{year}{2019}).

\bibitem[{\citenamefont{Norcia et~al.}(2018)\citenamefont{Norcia, Young, and
  Kaufman}}]{norcia2018microscopic}
\bibinfo{author}{\bibfnamefont{M.}~\bibnamefont{Norcia}},
  \bibinfo{author}{\bibfnamefont{A.}~\bibnamefont{Young}}, \bibnamefont{and}
  \bibinfo{author}{\bibfnamefont{A.}~\bibnamefont{Kaufman}},
  \bibinfo{journal}{Physical Review X} \textbf{\bibinfo{volume}{8}},
  \bibinfo{pages}{041054} (\bibinfo{year}{2018}).

\bibitem[{\citenamefont{Cooper et~al.}(2018)\citenamefont{Cooper, Covey,
  Madjarov, Porsev, Safronova, and Endres}}]{cooper2018alkaline}
\bibinfo{author}{\bibfnamefont{A.}~\bibnamefont{Cooper}},
  \bibinfo{author}{\bibfnamefont{J.~P.} \bibnamefont{Covey}},
  \bibinfo{author}{\bibfnamefont{I.~S.} \bibnamefont{Madjarov}},
  \bibinfo{author}{\bibfnamefont{S.~G.} \bibnamefont{Porsev}},
  \bibinfo{author}{\bibfnamefont{M.~S.} \bibnamefont{Safronova}},
  \bibnamefont{and} \bibinfo{author}{\bibfnamefont{M.}~\bibnamefont{Endres}},
  \bibinfo{journal}{Physical Review X} \textbf{\bibinfo{volume}{8}},
  \bibinfo{pages}{041055} (\bibinfo{year}{2018}).

\bibitem[{\citenamefont{Brennen et~al.}(1999)\citenamefont{Brennen, Caves,
  Jessen, and Deutsch}}]{brennen1999quantum}
\bibinfo{author}{\bibfnamefont{G.~K.} \bibnamefont{Brennen}},
  \bibinfo{author}{\bibfnamefont{C.~M.} \bibnamefont{Caves}},
  \bibinfo{author}{\bibfnamefont{P.~S.} \bibnamefont{Jessen}},
  \bibnamefont{and} \bibinfo{author}{\bibfnamefont{I.~H.}
  \bibnamefont{Deutsch}}, \bibinfo{journal}{Physical Review Letters}
  \textbf{\bibinfo{volume}{82}}, \bibinfo{pages}{1060} (\bibinfo{year}{1999}).

\bibitem[{\citenamefont{DiVincenzo}(2000)}]{divincenzo2000physical}
\bibinfo{author}{\bibfnamefont{D.~P.} \bibnamefont{DiVincenzo}},
  \bibinfo{journal}{Fortschritte der Physik: Progress of Physics}
  \textbf{\bibinfo{volume}{48}}, \bibinfo{pages}{771} (\bibinfo{year}{2000}).

\bibitem[{\citenamefont{Jaksch et~al.}(2000)\citenamefont{Jaksch, Cirac,
  Zoller, Rolston, C{\^o}t{\'e}, and Lukin}}]{jaksch2000fast}
\bibinfo{author}{\bibfnamefont{D.}~\bibnamefont{Jaksch}},
  \bibinfo{author}{\bibfnamefont{J.~I.} \bibnamefont{Cirac}},
  \bibinfo{author}{\bibfnamefont{P.}~\bibnamefont{Zoller}},
  \bibinfo{author}{\bibfnamefont{S.~L.} \bibnamefont{Rolston}},
  \bibinfo{author}{\bibfnamefont{R.}~\bibnamefont{C{\^o}t{\'e}}},
  \bibnamefont{and} \bibinfo{author}{\bibfnamefont{M.~D.} \bibnamefont{Lukin}},
  \bibinfo{journal}{Physical Review Letters} \textbf{\bibinfo{volume}{85}},
  \bibinfo{pages}{2208} (\bibinfo{year}{2000}).

\bibitem[{\citenamefont{Lukin et~al.}(2001)\citenamefont{Lukin, Fleischhauer,
  Cote, Duan, Jaksch, Cirac, and Zoller}}]{lukin2001dipole}
\bibinfo{author}{\bibfnamefont{M.}~\bibnamefont{Lukin}},
  \bibinfo{author}{\bibfnamefont{M.}~\bibnamefont{Fleischhauer}},
  \bibinfo{author}{\bibfnamefont{R.}~\bibnamefont{Cote}},
  \bibinfo{author}{\bibfnamefont{L.}~\bibnamefont{Duan}},
  \bibinfo{author}{\bibfnamefont{D.}~\bibnamefont{Jaksch}},
  \bibinfo{author}{\bibfnamefont{J.}~\bibnamefont{Cirac}}, \bibnamefont{and}
  \bibinfo{author}{\bibfnamefont{P.}~\bibnamefont{Zoller}},
  \bibinfo{journal}{Physical review letters} \textbf{\bibinfo{volume}{87}},
  \bibinfo{pages}{037901} (\bibinfo{year}{2001}).

\bibitem[{\citenamefont{Daley et~al.}(2008)\citenamefont{Daley, Boyd, Ye, and
  Zoller}}]{daley2008quantum}
\bibinfo{author}{\bibfnamefont{A.~J.} \bibnamefont{Daley}},
  \bibinfo{author}{\bibfnamefont{M.~M.} \bibnamefont{Boyd}},
  \bibinfo{author}{\bibfnamefont{J.}~\bibnamefont{Ye}}, \bibnamefont{and}
  \bibinfo{author}{\bibfnamefont{P.}~\bibnamefont{Zoller}},
  \bibinfo{journal}{Physical review letters} \textbf{\bibinfo{volume}{101}},
  \bibinfo{pages}{170504} (\bibinfo{year}{2008}).

\bibitem[{\citenamefont{Negretti et~al.}(2011)\citenamefont{Negretti,
  Treutlein, and Calarco}}]{negretti2011quantum}
\bibinfo{author}{\bibfnamefont{A.}~\bibnamefont{Negretti}},
  \bibinfo{author}{\bibfnamefont{P.}~\bibnamefont{Treutlein}},
  \bibnamefont{and} \bibinfo{author}{\bibfnamefont{T.}~\bibnamefont{Calarco}},
  \bibinfo{journal}{Quantum information processing}
  \textbf{\bibinfo{volume}{10}}, \bibinfo{pages}{721} (\bibinfo{year}{2011}).

\bibitem[{\citenamefont{Weitenberg
  et~al.}(2011{\natexlab{b}})\citenamefont{Weitenberg, Kuhr, M{\o}lmer, and
  Sherson}}]{weitenberg2011quantum}
\bibinfo{author}{\bibfnamefont{C.}~\bibnamefont{Weitenberg}},
  \bibinfo{author}{\bibfnamefont{S.}~\bibnamefont{Kuhr}},
  \bibinfo{author}{\bibfnamefont{K.}~\bibnamefont{M{\o}lmer}},
  \bibnamefont{and} \bibinfo{author}{\bibfnamefont{J.~F.}
  \bibnamefont{Sherson}}, \bibinfo{journal}{Physical Review A}
  \textbf{\bibinfo{volume}{84}}, \bibinfo{pages}{032322}
  (\bibinfo{year}{2011}{\natexlab{b}}).

\bibitem[{\citenamefont{Schneider and Saenz}(2012)}]{schneider2012quantum}
\bibinfo{author}{\bibfnamefont{P.-I.} \bibnamefont{Schneider}}
  \bibnamefont{and} \bibinfo{author}{\bibfnamefont{A.}~\bibnamefont{Saenz}},
  \bibinfo{journal}{Physical Review A} \textbf{\bibinfo{volume}{85}},
  \bibinfo{pages}{050304} (\bibinfo{year}{2012}).

\bibitem[{\citenamefont{J\o{}rgensen et~al.}(2014)\citenamefont{J\o{}rgensen,
  Bason, and Sherson}}]{byg2014Superlattice}
\bibinfo{author}{\bibfnamefont{N.~B.} \bibnamefont{J\o{}rgensen}},
  \bibinfo{author}{\bibfnamefont{M.~G.} \bibnamefont{Bason}}, \bibnamefont{and}
  \bibinfo{author}{\bibfnamefont{J.~F.} \bibnamefont{Sherson}},
  \bibinfo{journal}{Phys. Rev. A} \textbf{\bibinfo{volume}{89}},
  \bibinfo{pages}{032306} (\bibinfo{year}{2014}).

\bibitem[{\citenamefont{Pagano et~al.}(2019)\citenamefont{Pagano, Scazza, and
  Foss-Feig}}]{pagano2019fast}
\bibinfo{author}{\bibfnamefont{G.}~\bibnamefont{Pagano}},
  \bibinfo{author}{\bibfnamefont{F.}~\bibnamefont{Scazza}}, \bibnamefont{and}
  \bibinfo{author}{\bibfnamefont{M.}~\bibnamefont{Foss-Feig}},
  \bibinfo{journal}{Advanced Quantum Technologies}
  \textbf{\bibinfo{volume}{2}}, \bibinfo{pages}{1800067}
  (\bibinfo{year}{2019}).

\bibitem[{\citenamefont{Xia et~al.}(2015)\citenamefont{Xia, Lichtman, Maller,
  Carr, Piotrowicz, Isenhower, and Saffman}}]{xia2015randomized}
\bibinfo{author}{\bibfnamefont{T.}~\bibnamefont{Xia}},
  \bibinfo{author}{\bibfnamefont{M.}~\bibnamefont{Lichtman}},
  \bibinfo{author}{\bibfnamefont{K.}~\bibnamefont{Maller}},
  \bibinfo{author}{\bibfnamefont{A.}~\bibnamefont{Carr}},
  \bibinfo{author}{\bibfnamefont{M.}~\bibnamefont{Piotrowicz}},
  \bibinfo{author}{\bibfnamefont{L.}~\bibnamefont{Isenhower}},
  \bibnamefont{and} \bibinfo{author}{\bibfnamefont{M.}~\bibnamefont{Saffman}},
  \bibinfo{journal}{Physical review letters} \textbf{\bibinfo{volume}{114}},
  \bibinfo{pages}{100503} (\bibinfo{year}{2015}).

\bibitem[{\citenamefont{Wang et~al.}(2016)\citenamefont{Wang, Kumar, Wu, and
  Weiss}}]{wang2016single}
\bibinfo{author}{\bibfnamefont{Y.}~\bibnamefont{Wang}},
  \bibinfo{author}{\bibfnamefont{A.}~\bibnamefont{Kumar}},
  \bibinfo{author}{\bibfnamefont{T.-Y.} \bibnamefont{Wu}}, \bibnamefont{and}
  \bibinfo{author}{\bibfnamefont{D.~S.} \bibnamefont{Weiss}},
  \bibinfo{journal}{Science} \textbf{\bibinfo{volume}{352}},
  \bibinfo{pages}{1562} (\bibinfo{year}{2016}).

\bibitem[{\citenamefont{Mandel et~al.}(2003)\citenamefont{Mandel, Greiner,
  Widera, Rom, H{\"a}nsch, and Bloch}}]{mandel2003controlled}
\bibinfo{author}{\bibfnamefont{O.}~\bibnamefont{Mandel}},
  \bibinfo{author}{\bibfnamefont{M.}~\bibnamefont{Greiner}},
  \bibinfo{author}{\bibfnamefont{A.}~\bibnamefont{Widera}},
  \bibinfo{author}{\bibfnamefont{T.}~\bibnamefont{Rom}},
  \bibinfo{author}{\bibfnamefont{T.~W.} \bibnamefont{H{\"a}nsch}},
  \bibnamefont{and} \bibinfo{author}{\bibfnamefont{I.}~\bibnamefont{Bloch}},
  \bibinfo{journal}{Nature} \textbf{\bibinfo{volume}{425}},
  \bibinfo{pages}{937} (\bibinfo{year}{2003}).

\bibitem[{\citenamefont{Anderlini et~al.}(2007)\citenamefont{Anderlini, Lee,
  Brown, Sebby-Strabley, Phillips, and Porto}}]{anderlini2007controlled}
\bibinfo{author}{\bibfnamefont{M.}~\bibnamefont{Anderlini}},
  \bibinfo{author}{\bibfnamefont{P.~J.} \bibnamefont{Lee}},
  \bibinfo{author}{\bibfnamefont{B.~L.} \bibnamefont{Brown}},
  \bibinfo{author}{\bibfnamefont{J.}~\bibnamefont{Sebby-Strabley}},
  \bibinfo{author}{\bibfnamefont{W.~D.} \bibnamefont{Phillips}},
  \bibnamefont{and} \bibinfo{author}{\bibfnamefont{J.}~\bibnamefont{Porto}},
  \bibinfo{journal}{Nature} \textbf{\bibinfo{volume}{448}},
  \bibinfo{pages}{452} (\bibinfo{year}{2007}).

\bibitem[{\citenamefont{Wilk et~al.}(2010)\citenamefont{Wilk, Ga{\"e}tan,
  Evellin, Wolters, Miroshnychenko, Grangier, and
  Browaeys}}]{wilk2010entanglement}
\bibinfo{author}{\bibfnamefont{T.}~\bibnamefont{Wilk}},
  \bibinfo{author}{\bibfnamefont{A.}~\bibnamefont{Ga{\"e}tan}},
  \bibinfo{author}{\bibfnamefont{C.}~\bibnamefont{Evellin}},
  \bibinfo{author}{\bibfnamefont{J.}~\bibnamefont{Wolters}},
  \bibinfo{author}{\bibfnamefont{Y.}~\bibnamefont{Miroshnychenko}},
  \bibinfo{author}{\bibfnamefont{P.}~\bibnamefont{Grangier}}, \bibnamefont{and}
  \bibinfo{author}{\bibfnamefont{A.}~\bibnamefont{Browaeys}},
  \bibinfo{journal}{Physical Review Letters} \textbf{\bibinfo{volume}{104}},
  \bibinfo{pages}{010502} (\bibinfo{year}{2010}).

\bibitem[{\citenamefont{Zhang et~al.}(2010)\citenamefont{Zhang, Isenhower,
  Gill, Walker, and Saffman}}]{zhang2010deterministic}
\bibinfo{author}{\bibfnamefont{X.}~\bibnamefont{Zhang}},
  \bibinfo{author}{\bibfnamefont{L.}~\bibnamefont{Isenhower}},
  \bibinfo{author}{\bibfnamefont{A.}~\bibnamefont{Gill}},
  \bibinfo{author}{\bibfnamefont{T.}~\bibnamefont{Walker}}, \bibnamefont{and}
  \bibinfo{author}{\bibfnamefont{M.}~\bibnamefont{Saffman}},
  \bibinfo{journal}{Physical Review A} \textbf{\bibinfo{volume}{82}},
  \bibinfo{pages}{030306} (\bibinfo{year}{2010}).

\bibitem[{\citenamefont{Isenhower et~al.}(2010)\citenamefont{Isenhower, Urban,
  Zhang, Gill, Henage, Johnson, Walker, and
  Saffman}}]{isenhower2010demonstration}
\bibinfo{author}{\bibfnamefont{L.}~\bibnamefont{Isenhower}},
  \bibinfo{author}{\bibfnamefont{E.}~\bibnamefont{Urban}},
  \bibinfo{author}{\bibfnamefont{X.}~\bibnamefont{Zhang}},
  \bibinfo{author}{\bibfnamefont{A.}~\bibnamefont{Gill}},
  \bibinfo{author}{\bibfnamefont{T.}~\bibnamefont{Henage}},
  \bibinfo{author}{\bibfnamefont{T.~A.} \bibnamefont{Johnson}},
  \bibinfo{author}{\bibfnamefont{T.}~\bibnamefont{Walker}}, \bibnamefont{and}
  \bibinfo{author}{\bibfnamefont{M.}~\bibnamefont{Saffman}},
  \bibinfo{journal}{Physical review letters} \textbf{\bibinfo{volume}{104}},
  \bibinfo{pages}{010503} (\bibinfo{year}{2010}).

\bibitem[{\citenamefont{Maller et~al.}(2015)\citenamefont{Maller, Lichtman,
  Xia, Sun, Piotrowicz, Carr, Isenhower, and Saffman}}]{maller2015rydberg}
\bibinfo{author}{\bibfnamefont{K.}~\bibnamefont{Maller}},
  \bibinfo{author}{\bibfnamefont{M.}~\bibnamefont{Lichtman}},
  \bibinfo{author}{\bibfnamefont{T.}~\bibnamefont{Xia}},
  \bibinfo{author}{\bibfnamefont{Y.}~\bibnamefont{Sun}},
  \bibinfo{author}{\bibfnamefont{M.}~\bibnamefont{Piotrowicz}},
  \bibinfo{author}{\bibfnamefont{A.}~\bibnamefont{Carr}},
  \bibinfo{author}{\bibfnamefont{L.}~\bibnamefont{Isenhower}},
  \bibnamefont{and} \bibinfo{author}{\bibfnamefont{M.}~\bibnamefont{Saffman}},
  \bibinfo{journal}{Physical Review A} \textbf{\bibinfo{volume}{92}},
  \bibinfo{pages}{022336} (\bibinfo{year}{2015}).

\bibitem[{\citenamefont{Kaufman et~al.}(2015)\citenamefont{Kaufman, Lester,
  Foss-Feig, Wall, Rey, and Regal}}]{kaufman2015entangling}
\bibinfo{author}{\bibfnamefont{A.}~\bibnamefont{Kaufman}},
  \bibinfo{author}{\bibfnamefont{B.}~\bibnamefont{Lester}},
  \bibinfo{author}{\bibfnamefont{M.}~\bibnamefont{Foss-Feig}},
  \bibinfo{author}{\bibfnamefont{M.}~\bibnamefont{Wall}},
  \bibinfo{author}{\bibfnamefont{A.}~\bibnamefont{Rey}}, \bibnamefont{and}
  \bibinfo{author}{\bibfnamefont{C.}~\bibnamefont{Regal}},
  \bibinfo{journal}{Nature} \textbf{\bibinfo{volume}{527}},
  \bibinfo{pages}{208} (\bibinfo{year}{2015}).

\bibitem[{\citenamefont{Jau et~al.}(2016)\citenamefont{Jau, Hankin, Keating,
  Deutsch, and Biedermann}}]{jau2016entangling}
\bibinfo{author}{\bibfnamefont{Y.-Y.} \bibnamefont{Jau}},
  \bibinfo{author}{\bibfnamefont{A.}~\bibnamefont{Hankin}},
  \bibinfo{author}{\bibfnamefont{T.}~\bibnamefont{Keating}},
  \bibinfo{author}{\bibfnamefont{I.}~\bibnamefont{Deutsch}}, \bibnamefont{and}
  \bibinfo{author}{\bibfnamefont{G.}~\bibnamefont{Biedermann}},
  \bibinfo{journal}{Nature Physics} \textbf{\bibinfo{volume}{12}},
  \bibinfo{pages}{71} (\bibinfo{year}{2016}).

\bibitem[{\citenamefont{Levine et~al.}(2018)\citenamefont{Levine, Keesling,
  Omran, Bernien, Schwartz, Zibrov, Endres, Greiner, Vuleti{\'c}, and
  Lukin}}]{levine2018high}
\bibinfo{author}{\bibfnamefont{H.}~\bibnamefont{Levine}},
  \bibinfo{author}{\bibfnamefont{A.}~\bibnamefont{Keesling}},
  \bibinfo{author}{\bibfnamefont{A.}~\bibnamefont{Omran}},
  \bibinfo{author}{\bibfnamefont{H.}~\bibnamefont{Bernien}},
  \bibinfo{author}{\bibfnamefont{S.}~\bibnamefont{Schwartz}},
  \bibinfo{author}{\bibfnamefont{A.~S.} \bibnamefont{Zibrov}},
  \bibinfo{author}{\bibfnamefont{M.}~\bibnamefont{Endres}},
  \bibinfo{author}{\bibfnamefont{M.}~\bibnamefont{Greiner}},
  \bibinfo{author}{\bibfnamefont{V.}~\bibnamefont{Vuleti{\'c}}},
  \bibnamefont{and} \bibinfo{author}{\bibfnamefont{M.~D.} \bibnamefont{Lukin}},
  \bibinfo{journal}{Physical review letters} \textbf{\bibinfo{volume}{121}},
  \bibinfo{pages}{123603} (\bibinfo{year}{2018}).

\bibitem[{\citenamefont{Browaeys et~al.}(2016)\citenamefont{Browaeys, Barredo,
  and Lahaye}}]{browaeys2016experimental}
\bibinfo{author}{\bibfnamefont{A.}~\bibnamefont{Browaeys}},
  \bibinfo{author}{\bibfnamefont{D.}~\bibnamefont{Barredo}}, \bibnamefont{and}
  \bibinfo{author}{\bibfnamefont{T.}~\bibnamefont{Lahaye}},
  \bibinfo{journal}{Journal of Physics B: Atomic, Molecular and Optical
  Physics} \textbf{\bibinfo{volume}{49}}, \bibinfo{pages}{152001}
  (\bibinfo{year}{2016}).

\bibitem[{\citenamefont{Saffman et~al.}(2010)\citenamefont{Saffman, Walker, and
  M{\o}lmer}}]{saffman2010quantum}
\bibinfo{author}{\bibfnamefont{M.}~\bibnamefont{Saffman}},
  \bibinfo{author}{\bibfnamefont{T.~G.} \bibnamefont{Walker}},
  \bibnamefont{and}
  \bibinfo{author}{\bibfnamefont{K.}~\bibnamefont{M{\o}lmer}},
  \bibinfo{journal}{Reviews of Modern Physics} \textbf{\bibinfo{volume}{82}},
  \bibinfo{pages}{2313} (\bibinfo{year}{2010}).

\bibitem[{\citenamefont{Saffman}(2016)}]{saffman2016quantum}
\bibinfo{author}{\bibfnamefont{M.}~\bibnamefont{Saffman}},
  \bibinfo{journal}{Journal of Physics B: Atomic, Molecular and Optical
  Physics} \textbf{\bibinfo{volume}{49}}, \bibinfo{pages}{202001}
  (\bibinfo{year}{2016}).

\bibitem[{\citenamefont{Jaksch et~al.}(1999)\citenamefont{Jaksch, Briegel,
  Cirac, Gardiner, and Zoller}}]{jaksch1999entanglement}
\bibinfo{author}{\bibfnamefont{D.}~\bibnamefont{Jaksch}},
  \bibinfo{author}{\bibfnamefont{H.-J.} \bibnamefont{Briegel}},
  \bibinfo{author}{\bibfnamefont{J.}~\bibnamefont{Cirac}},
  \bibinfo{author}{\bibfnamefont{C.}~\bibnamefont{Gardiner}}, \bibnamefont{and}
  \bibinfo{author}{\bibfnamefont{P.}~\bibnamefont{Zoller}},
  \bibinfo{journal}{Physical Review Letters} \textbf{\bibinfo{volume}{82}},
  \bibinfo{pages}{1975} (\bibinfo{year}{1999}).

\bibitem[{\citenamefont{Calarco et~al.}(2000)\citenamefont{Calarco, Hinds,
  Jaksch, Schmiedmayer, Cirac, and Zoller}}]{calarco2000quantum}
\bibinfo{author}{\bibfnamefont{T.}~\bibnamefont{Calarco}},
  \bibinfo{author}{\bibfnamefont{E.}~\bibnamefont{Hinds}},
  \bibinfo{author}{\bibfnamefont{D.}~\bibnamefont{Jaksch}},
  \bibinfo{author}{\bibfnamefont{J.}~\bibnamefont{Schmiedmayer}},
  \bibinfo{author}{\bibfnamefont{J.}~\bibnamefont{Cirac}}, \bibnamefont{and}
  \bibinfo{author}{\bibfnamefont{P.}~\bibnamefont{Zoller}},
  \bibinfo{journal}{Physical Review A} \textbf{\bibinfo{volume}{61}},
  \bibinfo{pages}{022304} (\bibinfo{year}{2000}).

\bibitem[{\citenamefont{Hayes et~al.}(2007)\citenamefont{Hayes, Julienne, and
  Deutsch}}]{hayes2007quantumlogic}
\bibinfo{author}{\bibfnamefont{D.}~\bibnamefont{Hayes}},
  \bibinfo{author}{\bibfnamefont{P.~S.} \bibnamefont{Julienne}},
  \bibnamefont{and} \bibinfo{author}{\bibfnamefont{I.~H.}
  \bibnamefont{Deutsch}}, \bibinfo{journal}{Phys. Rev. Lett.}
  \textbf{\bibinfo{volume}{98}}, \bibinfo{pages}{070501}
  (\bibinfo{year}{2007}).

\bibitem[{\citenamefont{De~Chiara et~al.}(2008)\citenamefont{De~Chiara,
  Calarco, Anderlini, Montangero, Lee, Brown, Phillips, and
  Porto}}]{de2008optimal}
\bibinfo{author}{\bibfnamefont{G.}~\bibnamefont{De~Chiara}},
  \bibinfo{author}{\bibfnamefont{T.}~\bibnamefont{Calarco}},
  \bibinfo{author}{\bibfnamefont{M.}~\bibnamefont{Anderlini}},
  \bibinfo{author}{\bibfnamefont{S.}~\bibnamefont{Montangero}},
  \bibinfo{author}{\bibfnamefont{P.}~\bibnamefont{Lee}},
  \bibinfo{author}{\bibfnamefont{B.}~\bibnamefont{Brown}},
  \bibinfo{author}{\bibfnamefont{W.}~\bibnamefont{Phillips}}, \bibnamefont{and}
  \bibinfo{author}{\bibfnamefont{J.}~\bibnamefont{Porto}},
  \bibinfo{journal}{Physical Review A} \textbf{\bibinfo{volume}{77}},
  \bibinfo{pages}{052333} (\bibinfo{year}{2008}).

\bibitem[{\citenamefont{Mundt and Tannor}(2009)}]{mundt2009optimalcontrol}
\bibinfo{author}{\bibfnamefont{M.}~\bibnamefont{Mundt}} \bibnamefont{and}
  \bibinfo{author}{\bibfnamefont{D.~J.} \bibnamefont{Tannor}},
  \bibinfo{journal}{New Journal of Physics} \textbf{\bibinfo{volume}{11}},
  \bibinfo{pages}{105038} (\bibinfo{year}{2009}).

\bibitem[{\citenamefont{Anderlini et~al.}(2006)\citenamefont{Anderlini,
  Sebby-Strabley, Kruse, Porto, and Phillips}}]{anderlini2006controlled}
\bibinfo{author}{\bibfnamefont{M.}~\bibnamefont{Anderlini}},
  \bibinfo{author}{\bibfnamefont{J.}~\bibnamefont{Sebby-Strabley}},
  \bibinfo{author}{\bibfnamefont{J.}~\bibnamefont{Kruse}},
  \bibinfo{author}{\bibfnamefont{J.~V.} \bibnamefont{Porto}}, \bibnamefont{and}
  \bibinfo{author}{\bibfnamefont{W.~D.} \bibnamefont{Phillips}},
  \bibinfo{journal}{Journal of Physics B: Atomic, Molecular and Optical
  Physics} \textbf{\bibinfo{volume}{39}}, \bibinfo{pages}{S199}
  (\bibinfo{year}{2006}).

\bibitem[{Note1()}]{Note1}
Note1, \bibinfo{note}{visit \protect \url
  {https://www.quatomic.com/quatomic_publications/} for animations of the
  single-particle densities.}

\bibitem[{\citenamefont{Treutlein et~al.}(2006)\citenamefont{Treutlein,
  H{\"a}nsch, Reichel, Negretti, Cirone, and Calarco}}]{treutlein2006microwave}
\bibinfo{author}{\bibfnamefont{P.}~\bibnamefont{Treutlein}},
  \bibinfo{author}{\bibfnamefont{T.~W.} \bibnamefont{H{\"a}nsch}},
  \bibinfo{author}{\bibfnamefont{J.}~\bibnamefont{Reichel}},
  \bibinfo{author}{\bibfnamefont{A.}~\bibnamefont{Negretti}},
  \bibinfo{author}{\bibfnamefont{M.~A.} \bibnamefont{Cirone}},
  \bibnamefont{and} \bibinfo{author}{\bibfnamefont{T.}~\bibnamefont{Calarco}},
  \bibinfo{journal}{Physical Review A} \textbf{\bibinfo{volume}{74}},
  \bibinfo{pages}{022312} (\bibinfo{year}{2006}).

\bibitem[{\citenamefont{Olshanii}(1998)}]{olshanii}
\bibinfo{author}{\bibfnamefont{M.}~\bibnamefont{Olshanii}},
  \bibinfo{journal}{Physical Review Letters} \textbf{\bibinfo{volume}{81}},
  \bibinfo{pages}{938} (\bibinfo{year}{1998}).

\bibitem[{\citenamefont{S{\o}rensen et~al.}(2019)\citenamefont{S{\o}rensen,
  Jensen, Heinzel, and Sherson}}]{sorensen2019qengine}
\bibinfo{author}{\bibfnamefont{J.}~\bibnamefont{S{\o}rensen}},
  \bibinfo{author}{\bibfnamefont{J.}~\bibnamefont{Jensen}},
  \bibinfo{author}{\bibfnamefont{T.}~\bibnamefont{Heinzel}}, \bibnamefont{and}
  \bibinfo{author}{\bibfnamefont{J.}~\bibnamefont{Sherson}},
  \bibinfo{journal}{Computer Physics Communications}
  \textbf{\bibinfo{volume}{243}}, \bibinfo{pages}{135} (\bibinfo{year}{2019}).

\bibitem[{\citenamefont{Julienne et~al.}(1997)\citenamefont{Julienne, Mies,
  Tiesinga, and Williams}}]{julienne1997collisional}
\bibinfo{author}{\bibfnamefont{P.~S.} \bibnamefont{Julienne}},
  \bibinfo{author}{\bibfnamefont{F.}~\bibnamefont{Mies}},
  \bibinfo{author}{\bibfnamefont{E.}~\bibnamefont{Tiesinga}}, \bibnamefont{and}
  \bibinfo{author}{\bibfnamefont{C.~J.} \bibnamefont{Williams}},
  \bibinfo{journal}{Physical review letters} \textbf{\bibinfo{volume}{78}},
  \bibinfo{pages}{1880} (\bibinfo{year}{1997}).

\end{thebibliography}
\end{document}